\documentclass[journal]{IEEEtran}
\ifCLASSINFOpdf
\else
   \usepackage{graphicx}
   \graphicspath{{../eps/}}
   \DeclareGraphicsExtensions{.eps}
\fi

\usepackage{amsmath}
\usepackage{amsfonts,amssymb}
\usepackage{amssymb}
\usepackage{wrapfig}
\usepackage{psfrag}
\usepackage{epstopdf}
\usepackage{cite}
\usepackage{graphicx}
\usepackage{subfigure}
\usepackage{threeparttable}
\usepackage{cases}
\usepackage{subeqnarray}
\usepackage{color}
\usepackage{underscore}
\usepackage{verbatim}
\usepackage{bm}
\usepackage{stfloats}
\usepackage{xpatch}
\newtheorem{theorem}{Theorem}

\newtheorem{algorithm}{Algorithm}

\usepackage{algorithm}
\usepackage{algorithmic}

\makeatletter
\begin{document}
\title{Delay Alignment Modulation: Manipulating Channel Delay Spread for Efficient Single- \\ and Multi-Carrier Communication}

%
%
%

\author{Haiquan~Lu
        and
        Yong~Zeng,~\IEEEmembership{Senior Member,~IEEE}
\thanks{This work was supported by the National Key R\&D Program of China with Grant number 2019YFB1803400, and also by the Natural Science Foundation of China under Grant 62071114. (\emph{Corresponding author: Yong Zeng.}) }
\thanks{Haiquan Lu and Yong Zeng are with the National Mobile Communications Research Laboratory, Southeast University, Nanjing 210096, China, and are also with the Purple Mountain Laboratories, Nanjing 211111, China (e-mail: \{haiquanlu, yong_zeng\}@seu.edu.cn). }
}

\maketitle
\begin{abstract}
The evolution of mobile communication networks has always been accompanied by the advancement of inter-symbol interference (ISI) mitigation techniques, from equalization in the second-generation (2G), spread spectrum and RAKE receiver in the third generation (3G), to orthogonal frequency-division multiplexing (OFDM) in the fourth-generation (4G) and fifth-generation (5G). Looking forward towards the sixth-generation (6G), by exploiting the high spatial resolution brought by large antenna arrays and the multi-path sparsity of millimeter wave (mmWave) and Terahertz channels, a novel ISI mitigation technique termed \emph{delay alignment modulation} (DAM) was recently proposed. However, existing works only consider the single-carrier perfect DAM, which is feasible only when the number of base station (BS) antennas is no smaller than that of channel paths, so that all multi-path signal components can be aligned for arriving at the receiver simultaneously and constructively. This imposes stringent requirements on the number of BS antennas and multi-path sparsity. In this paper, we propose a generic DAM technique to manipulate the channel delay spread via spatial-delay processing, thus providing a flexible framework to combat channel time dispersion for efficient single- or multi-carrier transmissions. To gain some insights, we first show that when the number of BS antennas is much larger than that of channel paths, perfect delay alignment can be achieved to transform the time-dispersive channel to time non-dispersive channel with the simple delay pre-compensation and path-based maximal-ratio transmission (MRT) beamforming. When perfect DAM is infeasible or undesirable, the proposed generic DAM technique can be applied to significantly reduce the channel delay spread. Based on such results, we further propose the novel DAM-OFDM technique, which is able to save the cyclic prefix (CP) overhead or mitigate the peak-to-average-power ratio (PAPR) issue suffered by conventional OFDM. We show that the proposed DAM-OFDM involves joint frequency- and time-domain beamforming optimization, for which a closed-form solution is derived. Simulation results show that the proposed DAM-OFDM achieves significant performance gains over the conventional OFDM, in terms of spectral efficiency, bit error rate (BER) and PAPR.
\end{abstract}


\IEEEpeerreviewmaketitle
\section{Introduction}
 Multi-path signal propagation is one of the intrinsic properties of wireless communications, due to signal reflection, diffraction, and scattering. It has been well understood that when the symbol duration of the signal is much larger than the multi-path delay spread, the main effect of multi-path propagation is fading -- the variation of signal amplitude due to the constructive or destructive superposition of the different signal copies arriving at the receiver. However, if the symbol duration is comparable or even smaller than the multi-path delay spread, the detrimental inter-symbol interference (ISI) will become a major impairment \cite{heath2018foundations,goldsmith2005wireless}. The delay spread of multi-path channels can be quantified in various ways, for which the simplest one is the time difference between the last and first arrival significant signal components \cite{goldsmith2005wireless,tse2005fundamentals}. Other commonly used metrics include the root mean square (RMS) delay spread and the average delay spread, which requires the information of the power delay profile (PDP) \cite{goldsmith2005wireless,varela2001rms}.

 Looking back over the past few decades, the evolution of mobile communication networks has always been accompanied by the advancement of ISI mitigation techniques. For example, for the second-generation (2G) mobile communication networks, time-domain equalization (TEQ) at the receiver has been applied to Global System for Mobile Communications (GSM) \cite{buljore1994theoretical,proakis1991adaptive,steele1992mobile}. When the channel state information (CSI) is available at the transmitter, pre-equalization can also be applied at the transmitter to address the ISI issue \cite{xia1997new,fischer2005precoding}. Both equalization and pre-equalization are usually achieved by finite impulse response (FIR) filter, whose number of taps increases with the channel delay spread. In the third generation (3G) era, spread spectrum and RAKE receiver were the main techniques for ISI mitigation \cite{goldsmith2005wireless,tse2005fundamentals}. The key idea of spread spectrum is to use signal bandwidth much larger than that is needed for data transmission, so that spreading code with desired auto- and cross-correlation properties can be used for signal modulation and demodulation. Together with RAKE receiver, spread spectrum is able to distinguish multi-path signal components in the temporal domain and enable their coherent combining. However, implementing RAKE receiver with all resolvable multi-paths incurs prohibitive complexity, and selective RAKE receiver with a small number of fingers is usually employed \cite{win1999virtual}.

 For the fourth-generation (4G) and fifth-generation (5G) mobile communication systems, multi-carrier modulation or its efficient digital implementation via orthogonal frequency-division multiplexing (OFDM) has been the dominant technology for ISI mitigation. OFDM converts high-speed serial data stream into multiple low-speed parallel data streams, so that the OFDM symbol duration is much larger than the channel delay spread \cite{goldsmith2005wireless,heath2018foundations}, which thus resolves the ISI issue. However, to avoid inter-block interference, a cyclic prefix (CP) with length no smaller than the channel delay spread should be inserted at each OFDM symbol. Therefore, for severe time-dispersive channels, a large number of sub-carriers are needed in order to maintain acceptable CP overhead. However, it is widely known that as more sub-carriers are used, OFDM suffers from various drawbacks such as high peak-to-average-power ratio (PAPR) \cite{han2005overview,hung2014papr,jones1994block}, vulnerability to carrier frequency offset (CFO) \cite{sathananthan2001probability,yao2005blind}, and severe out-of-band (OOB) emission \cite{farhang2011ofdm,nissel2017filter}. Numerous efforts have been devoted to addressing such issues. For example, amplitude clipping, single-carrier frequency-division multiple access (SC-FDMA), and vector OFDM are possible methods for PAPR reduction \cite{han2005overview,xia2001precoded}. To reduce the OOB emission of OFDM, the windowing or filter bank multi-carrier (FBMC) modulation have been extensively studied \cite{nissel2017filter}. More recently, to address the severe CFO issue in high-mobility scenarios, a new multi-carrier technique termed \emph{orthogonal time frequency space} (OTFS) modulation has received significant research attentions \cite{hadani2017orthogonal}. However, such existing techniques either incur performance loss or require complicated signal processing. For instance, amplitude clipping inevitably causes distortion to OFDM signals, while FBMC requires sophisticated filtering for signal transmission and reception. Besides, OTFS faces new challenges of high receiver complexity and signal processing latency due to the large dimensional OTFS equivalent channel matrix \cite{xiao2021overview}.

\begin{figure}[!t]
 \setlength{\belowcaptionskip}{-0.1cm}
 \centering
 \subfigure[Non-sparse multi-path channel]{
    \includegraphics[width=1.62in,height=0.74in]{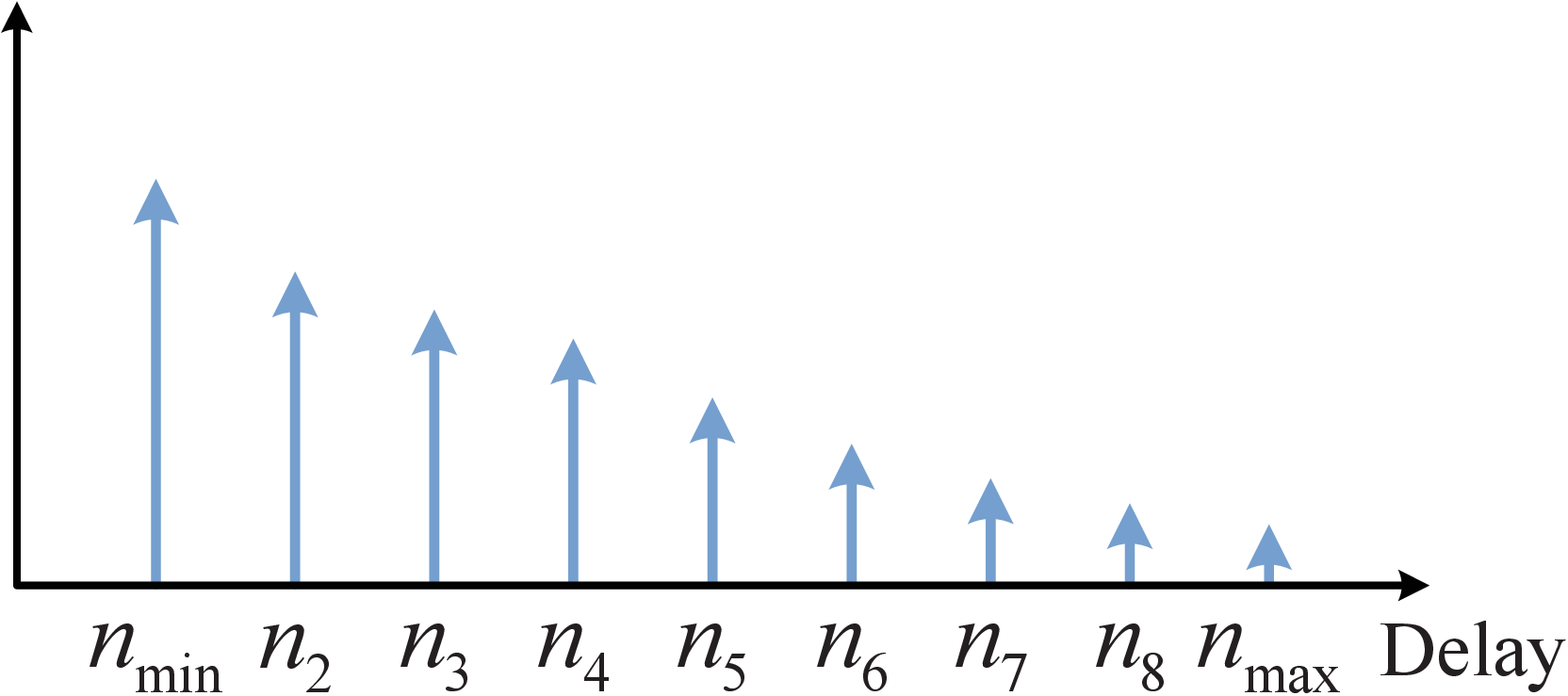}
 }
 \subfigure[Sparse multi-path channel]{
    \includegraphics[width=1.62in,height=0.74in]{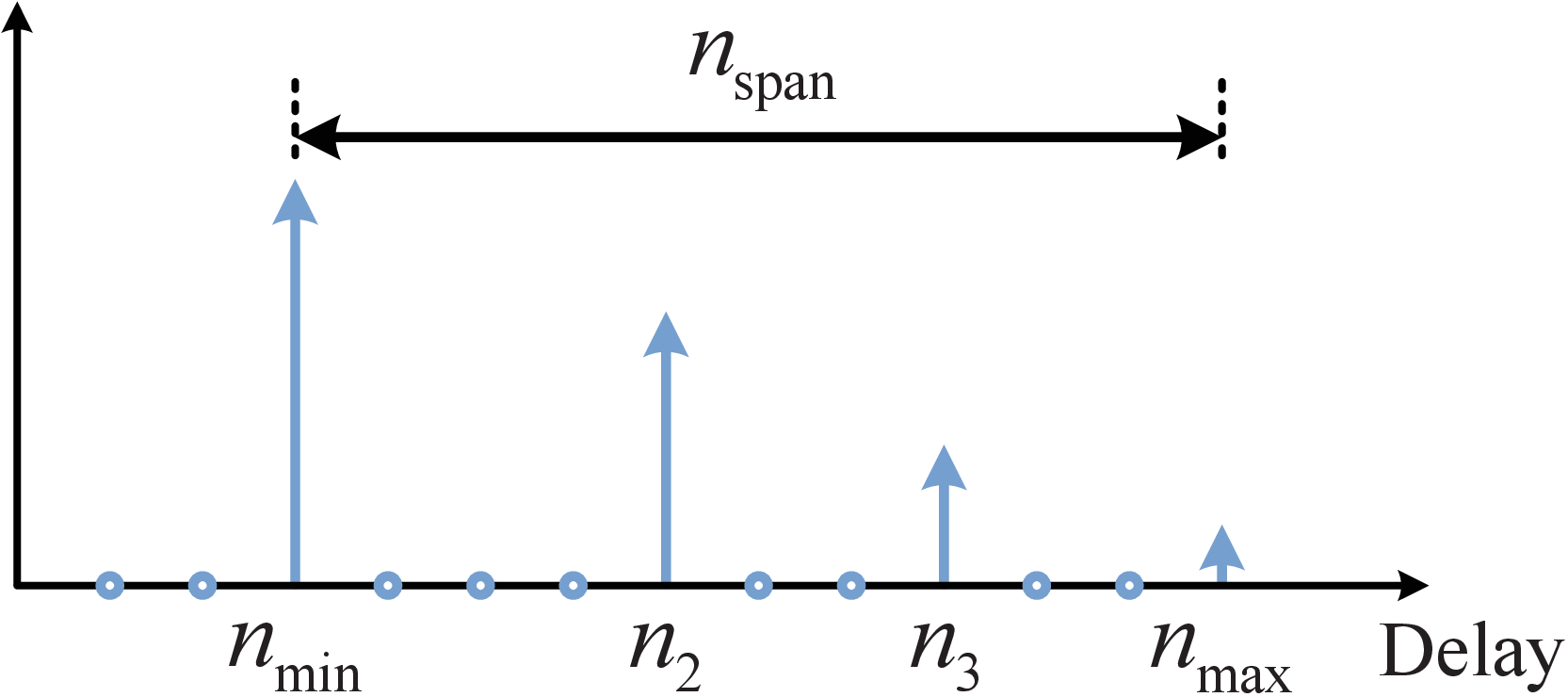}
 }
 \subfigure[Perfect DAM]{
    \includegraphics[width=1.62in,height=0.74in]{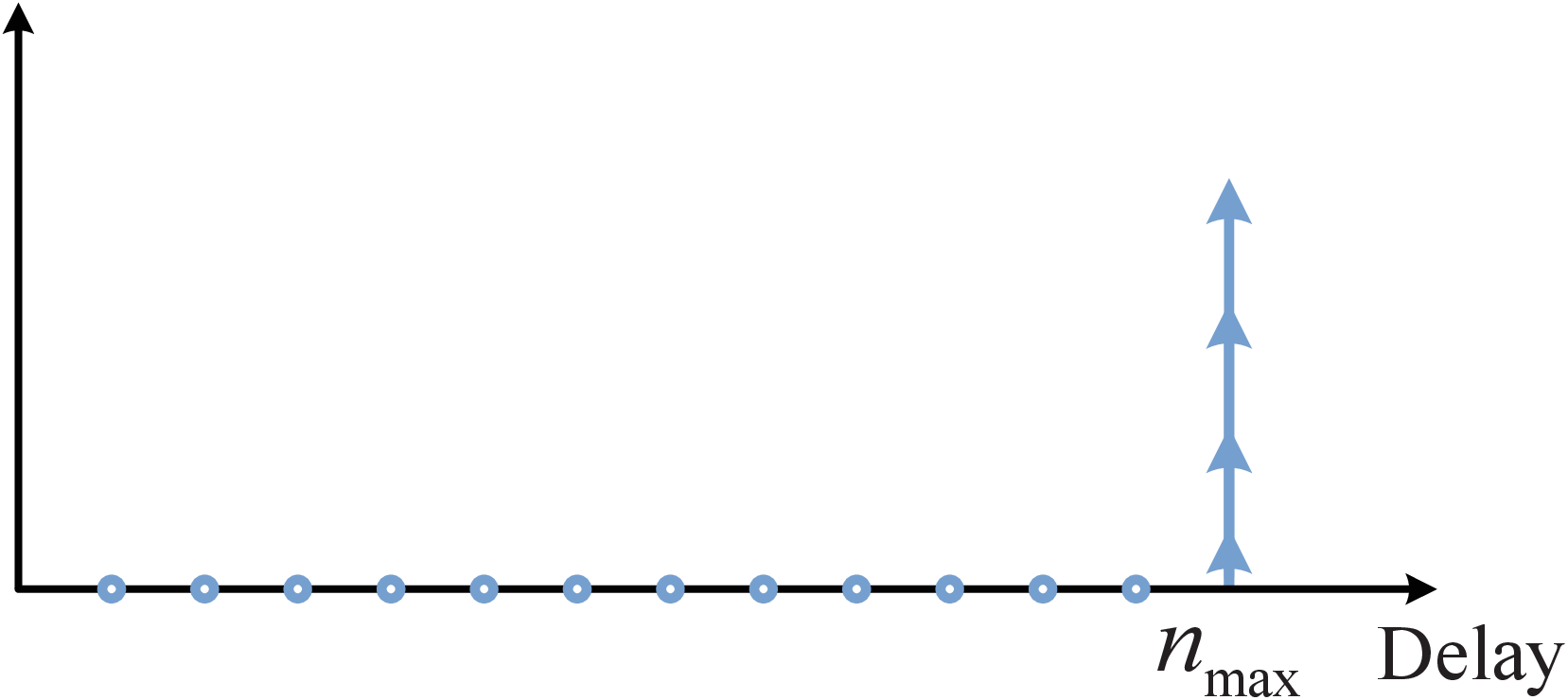}
 }
 \subfigure[Generic DAM]{
    \includegraphics[width=1.62in,height=0.75in]{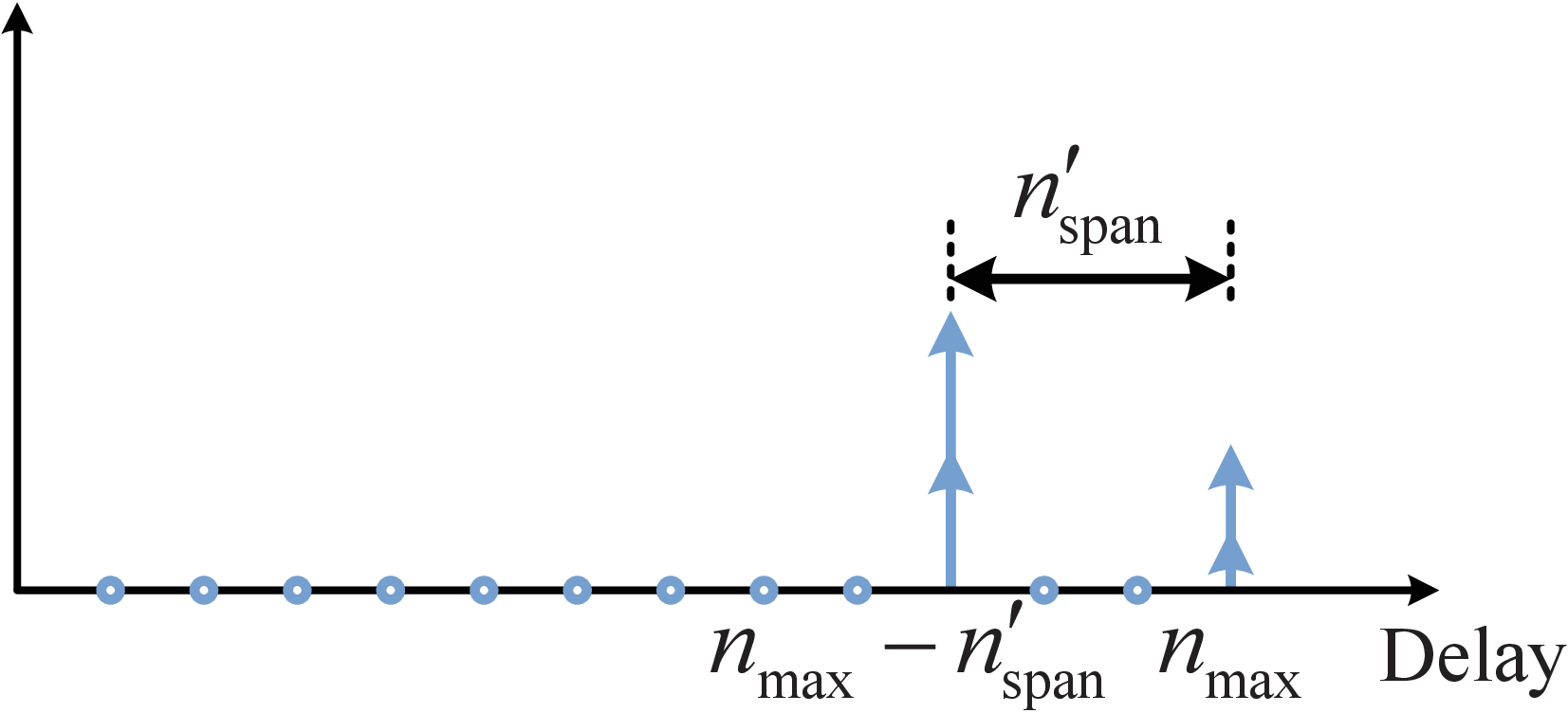}
 }
 \caption{Illustration of non-sparse and sparse multi-path channels, as well as the effect of DAM to manipulate the channel delay spread. With perfect DAM, all multi-paths are aligned to one single delay. With generic DAM, the multi-paths are aligned to a delay window with spread ${n'_{\rm{span}}} \ll {n_{\rm span}} = {n_{\max }} - {n_{\min }}$.}
 \label{AnIllustrationofMultipathSparsity}
 \end{figure}

 Looking forward towards the sixth-generation (6G) mobile communication systems, is it still possible to make further advancement for ISI mitigation?. Before answering this question, it is necessary to discuss several emerging trends for 6G that may offer new opportunities for ISI mitigation. Firstly, there is a general consensus that 6G will further exploit the high-frequency spectrum, such as millimeter wave (mmWave) and Terahertz signals to enable ultra high-capacity data transmission and super high-resolution sensing \cite{Latvaaho2019Keydrivers,dang2020should,you2021towards,zeng2016millimeter,zeng2018multi,huang2018wideband,rappaport2019wireless}. Compared to the sub-6G counterparts, one important feature of mmWave or Terahertz communications is that the multi-path channels are more sparse in the spatial and temporal domains \cite{rappaport2019wireless,xing2021millimeter,akdeniz2014millimeter,chen2021hybrid}, as illustrated in Fig.~\ref{AnIllustrationofMultipathSparsity}(a) and Fig.~\ref{AnIllustrationofMultipathSparsity}(b). Note that channel sparsity does not necessarily mean that the multi-path delay spread is small, since the last arrival significant path may still severely lag the first one, as illustrated in Fig.~\ref{AnIllustrationofMultipathSparsity}(b). Secondly, with massive multiple-input multiple-output (MIMO) being a reality in 5G, there has been fast-growing interest in further scaling up the antenna size significantly, leading to communication paradigm with extremely large-scale MIMO (XL-MIMO) \cite{lu2021communicating,cui2022channel,zhang2022beam}, ultra massive MIMO (UM-MIMO) \cite{WangJun2021General}, or extremely large aperture array (ELAA) \cite{bjornson2019massive}. This brings unprecedented spatial resolution to distinguish multi-path signals from/to different directions. Lastly, integrated localization, sensing and communication (ILSAC) will become a key feature for 6G \cite{tong20216g,xiao2022overview,zhang2021enabling}. In conjunction with mmWave/Terahertz communications and XL-MIMO, localization and sensing with extremely high resolution is expected to bring a paradigm shift for channel estimation, by extracting the features of each multi-path, such as its angle of arrival/departure (AoA/AoD) and propagation delay \cite{zhang2021enabling}, rather than only estimating the composite channels superimposed by all multi-paths in the time-frequency domain.

  The developing trends of 6G discussed above offer new opportunities to further advance the ISI mitigation techniques. By exploiting the high spatial resolution brought by large antenna arrays and the multi-path sparsity of mmWave and Terahertz channels, a novel ISI mitigation technique termed delay alignment modulation (DAM) was proposed in our recent work \cite{lu2022delay}. The key ideas of DAM are \emph{path delay pre/post-compensation} \cite{zeng2016millimeter,zeng2018multi} and \emph{path-based beamforming}. Specifically, with transmitter-side DAM, by deliberately introducing symbol delays at the transmitter to compensate for the respective multi-path delays of the channel, together with appropriate per-path-based beamforming, all multi-path signal components can be aligned for arriving at the receiver simultaneously and constructively, as illustrated in Fig.~\ref{AnIllustrationofMultipathSparsity}(c). The receiver-side DAM can be applied similarly via path delay post-compensation, which is analogous to RAKE receiver in the spatial domain, without having to use bandwidth that is much larger than the minimum necessary as in spread spectrum technique. Fundamentally, perfect DAM proposed in \cite{lu2022delay} is feasible only when the number of transmit antennas is no smaller than the number of multi-paths, and only single-carrier DAM is considered in \cite{lu2022delay}. In this paper, we propose a more generic DAM technique, rendering it possible to flexibly manipulate the channel delay spread for efficient single- or multi-carrier signal transmissions, while still benefiting from the multi-path signal components. Compared to perfect DAM, the generic DAM technique eases the requirement of antenna number and multi-path sparsity, without requiring that the number of transmit antennas is no smaller than that of channel paths. In particular, significantly reducing the channel delay spread makes it much easier to implement the equalization-based or RAKE receiver-based ISI mitigation techniques. On the other hand, for OFDM, reducing channel delay spread can save the CP overhead for a given number of sub-carriers, or mitigate the PAPR issue by using fewer sub-carriers without increasing the CP overhead. In fact, in the extreme case when the multi-path delays can be perfectly aligned, as illustrated in Fig.~\ref{AnIllustrationofMultipathSparsity}(c), a time non-dispersive channel is obtained, which enables equalization-free single-carrier transmission \cite{lu2022delay} or CP-free OFDM transmission, while still benefiting from the signal power contributed by all the multi-path components.

  It is worth remarking that DAM is significantly different from existing TEQ techniques, such as time-reversal (TR) \cite{emami2004matched,wang2011green,liu2012design,Why2016Chen,pitarokoilis2012optimality,aminjavaheri2017ofdm} and channel shortening \cite{melsa1996impulse,martin2005unification}. In particular, DAM takes full advantages of the high spatial resolution of large antenna arrays and multi-path sparsity of high-frequency signals to resolve the ISI issue with joint spatial-delay processing, rather than mainly relying on the time domain processing as in TR and channel shortening. For example, there have been extensive studies on TR technology \cite{han2016time,Why2016Chen} to create a virtual ``massive MIMO'' effect using only one single antenna or a few antennas, so as to reduce the deployment cost of standard massive MIMO systems \cite{han2016time,Why2016Chen}. However, such TR systems are unable to completely eliminate the ISI, and the rate back-off technique is needed, which significantly degrades the spectral efficiency \cite{emami2004matched,wang2011green}. On the other hand, though there are some studies combining massive MIMO with TR \cite{pitarokoilis2012optimality,pitaval2021channel}, they mostly leverage the multi-path channel as a matched filter, for which ISI can only be eliminated asymptotically as the number of the BS antennas goes to infinity \cite{pitarokoilis2012optimality,pitaval2021channel}. By contrast, DAM leverages the flexible path-based beamforming and delay pre-/post-compensation, so that ISI can be completely eliminated as long as the number of antennas is no smaller than the number of multi-paths. Furthermore, as presented in Section \ref{sectionDAMOFDM}, different from existing TEQ techniques, DAM enables the new paradigm of joint frequency- and time-domain beamforming, which unifies the multi-antenna single-carrier and multi-carrier transmission with complete ISI elimination. Joint time-frequency beamforming was also considered in \cite{li2005joint}, which aims to reduce the complexity of discrete Fourier transform (DFT) of the conventional MIMO-OFDM beamforming in the frequency-domain only, where one DFT block is needed for each antenna element. Thus, though both involving joint time-frequency beamforming, the work \cite{li2005joint} and the proposed DAM-OFDM in this paper consider two completely different problems. The main contributions of this paper are summarized as follows:
 \begin{itemize}[\IEEEsetlabelwidth{12)}]
 \item Firstly, for the asymptotic case when the number of transmit antennas $M_t$ is much larger than that of channel paths $L$, i.e., ${M_t} \gg L$, we show that the low-complexity per-path-based maximal-ratio transmission (MRT) beamforming, together with delay pre-compensation, is able to achieve perfect delay alignment. When ${M_t} \geq L$ but the multi-path channel vectors are non-orthogonal, we show that perfect delay alignment is still attainable with path-based zero-forcing (ZF) beamforming.
 \item Secondly, when perfect DAM is infeasible or undesirable, we propose the generic DAM technique to significantly reduce the channel delay spread, as illustrated in Fig.~\ref{AnIllustrationofMultipathSparsity}(d). This thus provides a flexible framework to combat time-dispersive channels for more efficient single- or multi-carrier transmissions.
 \item Lastly, to illustrate the benefits of reduced channel delay spread by DAM, we propose a novel DAM-OFDM scheme. DAM-OFDM includes the conventional OFDM and single-carrier transmissions as special cases, by flexibly varying the desired channel delay spread and the number of sub-carriers to save the CP overhead or mitigate the PAPR issue. We show that the beamforming design of DAM-OFDM involves the joint frequency- and time-domain beamforming, for which a closed-form solution is derived. Numerical results are provided to demonstrate the performance gain of DAM-OFDM over the conventional OFDM, in terms of spectral efficiency, bit error rate (BER), and PAPR.
\end{itemize}

 The rest of this paper is organized as follows. Section \ref{sectionSystemModelAndDAM} presents the system model and illustrates the key idea of perfect DAM. Section \ref{sectionManipulatingDelaySpread} proposes the generic DAM scheme to reduce the channel delay spread. In Section \ref{sectionDAMOFDM}, a novel communication scheme called DAM-OFDM is proposed. Simulation results are presented in Section \ref{sectionNumericalResults}. Finally, we conclude the paper in Section \ref{sectionConclusion}.

 \emph{Notations:} Scalars are denoted by italic letters. Vectors and matrices are denoted by bold-face lower- and upper-case letters, respectively. ${{\mathbb{C}}^{M \times N}}$ denotes the space of $M \times N$ complex-valued matrices. For a vector $\bf{x}$, $\left\| {\bf{x}} \right\|$ denotes its Euclidean norm. The notation $*$ denotes the linear convolution operation. The distribution of a circularly symmetric complex Gaussian (CSCG) random vector with mean $\bf{x}$ and covariance matrix $\bf{\Sigma}$ is denoted by ${\cal CN}\left( {\bf{x},\bf{{\Sigma}}} \right)$; and $\sim$ stands for ``distributed as". The symbol $j$ denotes the imaginary unit of complex numbers, with ${j^2} =  - 1$. For a real number $x$, $\left\lfloor x \right\rfloor $ denotes floor operations, and ${\rm round}\left( x \right)$ represents the nearest integer of $x$. ${\rm mod} \left( {x,y} \right)$ returns the remainder after division of $x$ by $y$. ${\mathbb E}\left({\cdot}\right)$ denotes the statistical expectation. For a set $\cal S$, $\left| {\cal S} \right|$ denotes its cardinality. ${\cal O}\left(\cdot  \right)$ denotes the standard big-O notation.

\section{System Model and Key Idea of DAM}\label{sectionSystemModelAndDAM}
\begin{figure}[!t]
 \centering
 \centerline{\includegraphics[width=3.3in,height=2.05in]{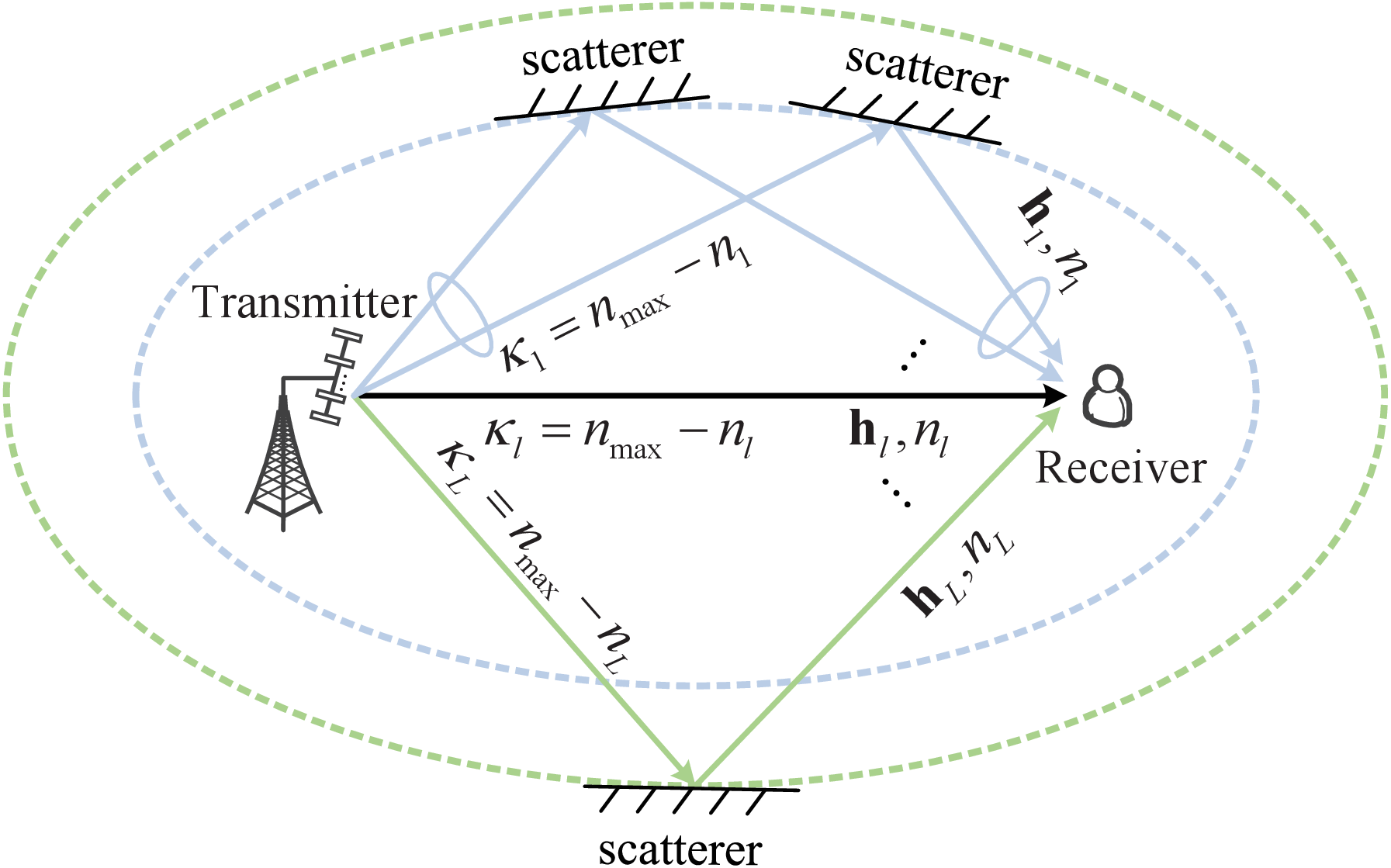}}
 \caption{A time-dispersive MISO channel, where scatterers located in the same ellipse contribute to multi-path components with a common temporal-resolvable delay but different AoDs \cite{lu2022delay}.}
 \label{systemModel}
 \end{figure}

 As shown in Fig.~\ref{systemModel}, we consider a time-dispersive multiple-input single-output (MISO) communication system similar to \cite{lu2022delay}, where the transmitter is equipped with $M_t$ antennas and the receiver has one antenna. For ease of exposition, we assume quasi-static block fading model, where the channel remains constant within each coherence block and may vary across different blocks. For each channel coherence block, the discrete-time equivalent of the channel impulse response can be expressed as{\footnote[1]{Within each channel coherence block, we have the time-invariant frequency-selective channel. For the more general time-variant frequency-selective channel, the DAM technique can be extended to \emph{delay-Doppler alignment modulation} (DDAM) to achieve Doppler-ISI double mitigation \cite{xiao2023exploiting,lu2023delayDoppler}.}}
 \begin{equation}\label{channelImpulseResponse}
 {\bf{h}}\left[ n \right] = \sum\limits_{l = 1}^L {{\bf{h}}_l\delta \left[ {n - {n_l}} \right]},
 \end{equation}
 where $L$ is the number of temporal-resolvable multi-paths with delay resolution ${1}/{B}$, with $B$ denoting the system bandwidth in hertz, ${\bf{h}}_l \in {{\mathbb{C}}^{{M_t} \times 1}}$ denotes the channel vector for the $l$th multi-path, and ${n_l} = {\rm{round}}\left( {{\tau _l}B} \right)$ denotes its discretized delay, with $\tau_l$ being the delay in second. Let ${n_{\min }} \triangleq \mathop {\min }\limits_{1 \le l \le L} {n_l}$ and ${n_{\max }} \triangleq \mathop {\max }\limits_{1 \le l \le L} {n_l}$ denote the minimum and maximum delay among all the $L$ multi-paths, respectively. Then the channel delay spread is defined as ${n_{\rm{span}}} = {n_{\max }} - {n_{\min }}$. To focus on the essential technique and reveal the fundamental performance limit of DAM, perfect CSI is assumed at the transmitter side. A preliminary effort on channel estimation and the performance characterization with CSI errors for DAM has been considered in our separate work \cite{ding2022channel}.

 Denote by ${\bf{x}}\left[ n \right] \in {{\mathbb{C}}^{{M_t} \times 1}}$ the discrete-time equivalent of the transmitted signal. Then the received signal is
 \begin{equation}\label{generalReceivedSignal}
 y\left[ n \right]  = {{\bf{h}}^H}\left[ n \right]*{\bf{x}}\left[ n \right] + z\left[ n \right]  = \sum\limits_{l = 1}^L {{\bf{h}}_l^H{\bf{x}}\left[ {n - {n_l}} \right]}  + z\left[ n \right],
 \end{equation}
 where $z\left[ n \right] \sim {\cal C}{\cal N}\left( {0,{\sigma ^2}} \right)$ is the additive white Gaussian noise (AWGN). Note that each temporal-resolvable multi-path may include several sub-paths that have a common delay but different AoDs, as illustrated in Fig.~\ref{systemModel}. Thus, ${\bf{h}}_l$ in \eqref{channelImpulseResponse} can be modelled as  ${{\bf{h}}_l} = {\alpha _l}\sum\nolimits_{i = 1}^{{\mu _l}} {{\upsilon _{li}}{\bf{a}}\left( {{\theta _{li}}} \right)}$, where ${\alpha _l}$ and $\mu_l$ denote the complex-valued path gain and the number of sub-paths for the $l$th multi-path, respectively, ${\upsilon _{li}}$ denotes the complex coefficient of the $i$th sub-path of path $l$, and $\theta_{li}$ and ${{\bf{a}}\left( {\theta _{li}} \right)} \in {{\mathbb{C}}^{{M_t} \times 1}}$ denote the AoD and the transmit array response vector of the $i$th sub-path of path $l$, respectively. Note that when certain AoD corresponds to more than one temporal-resolvable multi-paths, joint transmitter- and receiver-side DAM can be applied to resolve the multi-paths for MIMO systems.

 \begin{figure}[!t]
 \centering
 \centerline{\includegraphics[width=3.5in,height=2.625in]{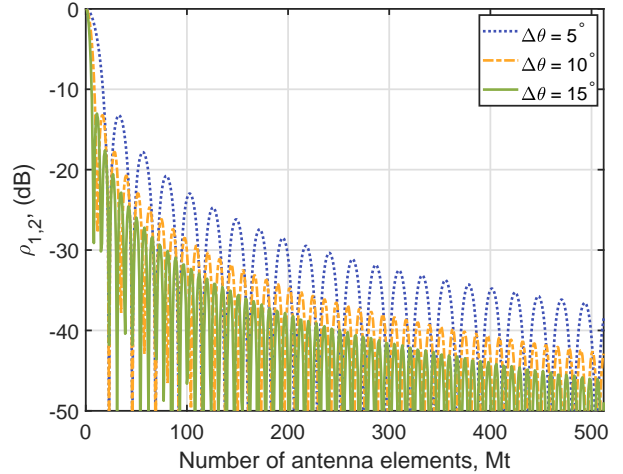}}
 \caption{The absolute square of the normalized inner product for channels associated with two path delays, where AoDs $\theta_1= 90^{\circ}$ and $\theta_2=\theta_1-\Delta \theta$.}
 \label{innerProductVersusAntennaNumber}
 \end{figure}
 Before presenting the key idea of our proposed DAM technique, we first study the correlation property of the MISO vectors $\{ {{\bf{h}}_l}\} _{l = 1}^L$ associated with different paths. For ease of illustration, we consider the basic uniform linear array (ULA) with adjacent elements separated by half wavelength, and the ULA aperture is $D = \left( {{M_t} - 1} \right)\lambda /2$, where $\lambda$ denotes the signal wavelength. When each temporal-resolvable multi-path component corresponds to one AoD only, i.e., ${\mu_l}=1$, $\forall l$, ${\bf h}_l$ can be modelled as \cite{tse2005fundamentals}
 \begin{equation}\label{arrayResponseVectorULA}
 {{\bf{h}}_l} = {\alpha _l}{\bf{a}}\left( {{\theta _l}} \right) = {\alpha _l}{\left[ {1,{e^{ - j\pi \cos {\theta _l}}}, \cdots ,{e^{ - j\pi \left( {{M_t} - 1} \right)\cos {\theta _l}}}} \right]^T}.
 \end{equation}
 Define ${\rho _{l,l'}} \triangleq  {\left| {{\bf{h}}_l^H{{\bf{h}}_{l'}}} \right|^2}/{\left\| {{{\bf{h}}_l}} \right\|^2}{\left\| {{{\bf{h}}_{l'}}} \right\|^2}$, which is the absolute square of the normalized inner product between channel vectors ${\bf h}_l$ and ${\bf h}_{l'}$. We have
 \begin{equation}\label{MISOVectorsInnerProduct}
 \begin{aligned}
 {\rho _{l,l'}} &= \frac{1}{{M_t^2}}{\left| {\sum\limits_{m = 1}^{{M_t}} {{e^{j\pi \left( {m - 1} \right)\left( {\cos {\theta _l} - \cos {\theta _{l'}}} \right)}}} } \right|^2} \\
 &= \frac{1}{{M_t^2}}{\left| {\frac{{\sin \left( {\frac{\pi }{2}{M_t}\left( {\cos {\theta _l} - \cos {\theta _{l'}}} \right)} \right)}}{{\sin \left( {\frac{\pi }{2}\left( {\cos {\theta _l} - \cos {\theta _{l'}}} \right)} \right)}}} \right|^2}.
 \end{aligned}
 \end{equation}
 It can be shown that as long as $\cos {\theta _l} \ne \cos {\theta _{l'}}$, as the number of antennas ${M_t} \to \infty $, we have ${\rho _{l,l'}} \to 0$. This is verified by Fig.~\ref{innerProductVersusAntennaNumber}, which plots $\rho_{1,2}$ versus $M_t$ with different AoD separations $\Delta \theta$, by letting ${\theta _1} = 90^{\circ}$ and $\theta_2=\theta_1-\Delta \theta$. It is observed that $\rho_{1,2}$ has the general trend of decreasing as $M_t$ increases, and the value is less than $-30$ dB when ${M_t} \ge 256$, even when the AoD separation is only $5^{\circ}$. This implies that as long as the temporal-resolvable multi-paths are associated with distinct AoDs, their corresponding  channel vectors become asymptotically orthogonal when ${M_t} \gg L$, i.e., ${\bf{h}}_l^H{{\bf{h}}_{l'}}/\left\| {{{\bf{h}}_l}} \right\|\left\| {{{\bf{h}}_{l'}}} \right\| \to 0$, $\forall l \ne l'$. This is analogous to the asymptotic orthogonality property for channels between different users in massive MIMO systems \cite{marzetta2010noncooperative}, which is extended to channels of the same user but different path delays. This thus motivates our proposed DAM technique, whose key idea is to apply path-based beamforming and delay pre-compensation at the transmitter to match their respective channel delays, as elaborated below.

 To gain some insights, we first consider single-carrier-based DAM transmission proposed in \cite{lu2022delay}, while the more general multi-carrier OFDM based DAM proposed in this paper will be presented in Section \ref{sectionDAMOFDM}. With single-carrier DAM, the transmitted signal is
 \begin{equation}\label{DAMTransmitSignal}
 {\bf{x}}\left[ n \right] = \sum\limits_{l = 1}^L {{{\bf{f}}_l}s\left[ {n - {\kappa _l}} \right]},
 \end{equation}
 where $s\left[ n \right]$ is the independent and identically distributed (i.i.d.) information-bearing symbols with normalized power ${\mathbb{E}}{\rm{[|}}s[n]{{\rm{|}}^2}] = 1$, ${{{\bf{f}}_l}} \in {{\mathbb{C}}^{{M_t} \times 1}}$ denotes the path-based transmit beamforming vector associated with path $l$, and $\kappa _l \ge 0$ is the deliberately introduced delay for the symbol sequence $s\left[ n \right]$ that aims to compensate for the delay of the $l$th channel path, with ${\kappa _l} \ne {\kappa _{l'}}$, $\forall l \ne l'$. The single-carrier DAM transmitter in \eqref{DAMTransmitSignal} can be easily implemented since delay pre-compensation by $\kappa_l$ simply means time shift of the sequence $s\left[ n \right]$, and the path-based transmit beamforming can be implemented similarly as standard beamforming techniques. The power of ${\bf{x}}\left[ n \right]$ in \eqref{DAMTransmitSignal} is ${\mathbb{E}}[ {{{\left\| {{\bf{x}}\left[ n \right]} \right\|}^2}}] = \sum\nolimits_{l = 1}^L {{{\left\| {{{\bf{f}}_l}} \right\|}^2}}  \le P$ \cite{lu2022delay}, where $P$ denotes the available transmit power. By letting ${\kappa _l} = {n_{\max }} - {n_l} \ge 0$, $\forall l$, and substituting  \eqref{DAMTransmitSignal} into \eqref{generalReceivedSignal}, the received signal for single-carrier DAM is
 \begin{equation}\label{DelayCompensationReceivedSignal}\
 \begin{aligned}
 y\left[ n \right]  = & \left( {\sum\limits_{l = 1}^L {{\bf{h}}_l^H{{\bf{f}}_l}} } \right)s\left[ {n - {n_{\max }}} \right]+ \\ &\sum\limits_{l = 1}^L {\sum\limits_{l' \ne l}^L {{\bf{h}}_l^H{{\bf{f}}_{l'}}s\left[ {n - {n_{\max }} + {n_{l'}} - {n_l}} \right]} }  + z\left[ n \right].
 \end{aligned}
 \end{equation}
 If the receiver is locked to the maximum delay $n_{\max}$, then the first term in \eqref{DelayCompensationReceivedSignal} contributes to the desired signal, while the second term is the ISI.

 However, when ${M_t} \gg L$, the channel vectors associated with different AoDs $\{{{\bf{h}}_l}\} _{l = 1}^L$ are asymptotically orthogonal, as illustrated in Fig.~\ref{innerProductVersusAntennaNumber}. Thus, by applying the simple path-based MRT beamforming, i.e., ${{\bf{f}}_l} = \sqrt P \xi {{\bf{h}}_l}$, $\forall l$, with $\xi  \triangleq 1/\sqrt {\sum\nolimits_{k = 1}^L {{{\left\| {{{\bf{h}}_k}} \right\|}^2}} }$, and scaling the received signal in \eqref{DelayCompensationReceivedSignal} by $\xi$, we have
 \begin{equation}\label{DelayCompensationReceivedSignalMRT}
 \begin{aligned}
 &\xi y\left[ n \right] = \sqrt P s\left[ {n - {n_{\max }}} \right] +\\
 &\sqrt P \sum\limits_{l = 1}^L {\sum\limits_{l' \ne l}^L {\frac{{{\bf{h}}_l^H{{\bf{h}}_{l'}}}}{{\sum\nolimits_{k = 1}^L {{{\left\| {{{\bf{h}}_k}} \right\|}^2}} }}s\left[ {n - {n_{\max }} + {n_{l'}} - {n_l}} \right]} }  + \xi z\left[ n \right].
 \end{aligned}
 \end{equation}
 When ${M_t} \gg L$, we have
 \begin{equation}
 \frac{{\left| {{\bf{h}}_l^H{{\bf{h}}_{l'}}} \right|}}{{\sum\nolimits_{k = 1}^L {{{\left\| {{{\bf{h}}_k}} \right\|}^2}} }} \le \frac{{\left| {{\bf{h}}_l^H{{\bf{h}}_{l'}}} \right|}}{{{{\left\| {{{\bf{h}}_l}} \right\|}^2} + {{\left\| {{{\bf{h}}_{l'}}} \right\|}^2}}} \le \frac{{\left| {{\bf{h}}_l^H{{\bf{h}}_{l'}}} \right|}}{{\left\| {{{\bf{h}}_l}} \right\|\left\| {{{\bf{h}}_{l'}}} \right\|}} \to 0, \ \forall l' \ne l,
 \end{equation}
 i.e., the ISI terms in \eqref{DelayCompensationReceivedSignalMRT} will approach to zero. By normalizing \eqref{DelayCompensationReceivedSignalMRT} with $\xi$ again, the signal in \eqref{DelayCompensationReceivedSignal} can be approximated as
 \begin{equation}\label{reducedDelayCompensationReceivedSignalMRT}
 y\left[ n \right] \approx \left( {\sqrt {P\sum\nolimits_{l = 1}^L {{{\left\| {{{\bf{h}}_l}} \right\|}^2}} } } \right)s\left[ {n - {n_{\max }}} \right] + z\left[ n \right].
 \end{equation}
 It is observed from \eqref{reducedDelayCompensationReceivedSignalMRT} that when ${M_t} \gg L$, with the simple delay pre-compensation and path-based MRT beamforming, the received signal is simply the symbol sequence $s\left[ n \right]$ delayed by one single delay $n_{\max}$ with a multiplicative gain contributed by all the $L$ multi-paths. As a result, the original time-dispersive channel has been transformed to  non-dispersive ISI-free AWGN channel, without applying sophisticated channel equalization or multi-carrier transmission. The received signal-to-noise ratio (SNR) of \eqref{reducedDelayCompensationReceivedSignalMRT} is $\gamma  = \bar P\sum\nolimits_{l = 1}^L {{{\left\| {{{\bf{h}}_l}} \right\|}^2}}$, where $\bar P \triangleq P/{\sigma ^2}$.

 On the other hand, when the per-path channels are non-orthogonal, provided that ${M_t} \ge L$, the ISI in \eqref{DelayCompensationReceivedSignal} can still be completely eliminated by path-based ZF beamforming. Specifically, $\{ {{\bf{f}}_l}\} _{l = 1}^L$ are designed so that
 \begin{equation}\label{DAMCondition}
 {\bf{h}}_l^H{{\bf{f}}_{l'}} = 0,\ \forall l \ne l'.
 \end{equation}
 The path-based ZF constraint in \eqref{DAMCondition} can be compactly written as ${\bf{H}}_{l'}^H{{\bf{f}}_{l'}} = {{\bf{0}}_{\left( {L - 1} \right) \times 1}}$, $\forall l' \in \left[1,L\right]$, where ${{\bf{H}}_{l'}} \in {{\mathbb C}^{{M_t} \times \left( {L - 1} \right)}} \triangleq \left[ {{{\bf{h}}_1}, \cdots ,{{\bf{h}}_{l' - 1}},{{\bf{h}}_{l' + 1}}, \cdots, {{\bf{h}}_L}} \right]$. The ZF conditions are feasible as long as ${\bf{H}}_{l'}^H$ has a non-empty null-space, which is true almost surely when $M_t \ge L$ and the AoDs of multi-path channels are random and uncorrelated. In this case, ${\bf{f}}_{l'}$ should lie in the orthogonal complement of ${{\bf{H}} _{l'}}$, which can be expressed as ${{\bf{f}}_{l'}} = {\bf{H}}_{l'}^ \bot {{\bf{b}}_{l'}}$, where ${\bf{H}}_{l'}^ \bot  \in {{\mathbb C}^{{M_t} \times {r_{l'}}}}$ denotes an orthonormal basis for the orthogonal complement of ${{\bf{H}} _{l'}}$, and ${{\bf{b}}_{l'}} \in {{\mathbb C}^{{r_{l'}} \times 1}}$ is the new beamforming vector to be designed, with ${r_{l'}} = {\rm{rank}}\left( {{\bf{H}}_{l'}^ \bot } \right)  = {M_t} - L + 1$. With path-based ZF beamforming, the received signal in \eqref{DelayCompensationReceivedSignal} reduces to
 \begin{equation}\label{DelayCompensationReceivedSignalEquivalent}
 y\left[ n \right] = \left( {\sum\limits_{l = 1}^L {{\bf{h}}_l^H{\bf{H}}_l^ \bot {{\bf{b}}_l}} } \right)s\left[ {n - {n_{\max }}} \right] + z\left[ n \right].
 \end{equation}

 \begin{figure}[!t]
 \centering
 \centerline{\includegraphics[width=3.3in,height=0.8in]{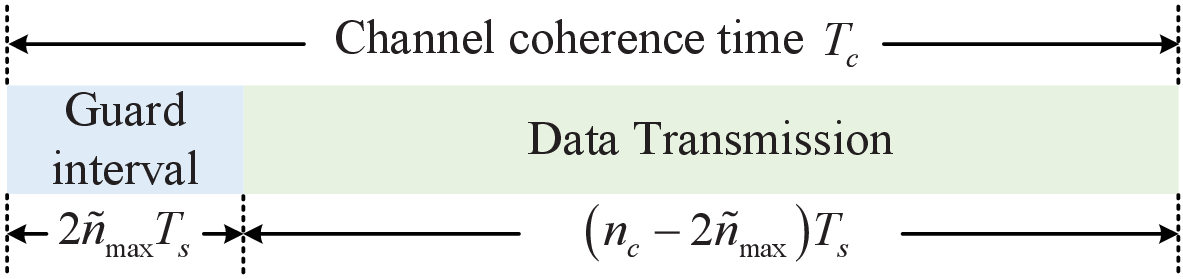}}
 \caption{An illustration of the single-carrier DAM block structure.}
 \label{DAMBlockStructure}
 \end{figure}

 As a result, with the simple DAM in \eqref{DAMTransmitSignal} and the path-based ZF beamforming in \eqref{DAMCondition}, perfect delay alignment and ISI-free single-carrier communication is still attainable \cite{lu2022delay}, without sophisticated equalization or multi-carrier transmission. The optimal path-based ZF beamforming to maximize the received SNR of \eqref{DelayCompensationReceivedSignalEquivalent} can be obtained by applying Cauchy-Schwarz inequality, given by
 \begin{equation}\label{optimalZFBeamformingSingleCarrier}
 {\bf{f}}_l^ \star  = \frac{{\sqrt P {\bf{H}}_l^ \bot {{\left( {{\bf{H}}_l^ \bot } \right)}^H}{{\bf{h}}_l}}}{{\sqrt {\sum\nolimits_{i = 1}^L {{{\| {{\bf{H}}_i^ \bot {{\left( {{\bf{H}}_i^ \bot } \right)}^H}{{\bf{h}}_i}} \|^2}}} } }},\ \forall l,
 \end{equation}
 and the resulting SNR is
 \begin{equation}\label{optimalSNRSingleCarrier}
 \gamma  = \bar P\sum\limits_{l = 1}^L {{{\left\| {{{\left( {{\bf{H}}_l^ \bot } \right)}^H}{{\bf{h}}_l}} \right\|}^2}}.
 \end{equation}
 It is worth mentioning that to avoid the ISI across consecutive channel coherence blocks, a guard interval of length $2{{\tilde n}_{\max }}$ can be inserted in the front of each channel coherence block, as illustrated in Fig.~\ref{DAMBlockStructure}, where ${{\tilde n}_{\max }}$ represents an upper bound of the maximum delay over all channel coherence blocks \cite{lu2022delay}, and ${n_c} = {T_c}/{T_s}$ represents the number of signal samples within each channel coherence time $T_c$, with ${T_s} = 1/B$.

\section{Manipulating Channel Delay Spread via DAM}\label{sectionManipulatingDelaySpread}
 The examples in the previous section demonstrate the capability of DAM to transform time-dispersive channels to non-dispersive channels, while still reaping the signal power contributed by all the multi-path components. However, perfect DAM targeting for zero delay spread may be infeasible (when $M_t <L$) or undesirable, since ZF beamforming to eliminate all ISI restricts the beamforming space for desired signal enhancement. Therefore, in this section, we present the more generic DAM technique, whose target is to reduce the channel delay spread to a certain desired value. As such, more efficient single- or multi-carrier signal transmissions can be applied to the resulting effective channel with reduced delay spread.
 \begin{figure}[!t]
 \centering
 \centerline{\includegraphics[width=3.55in,height=1.65in]{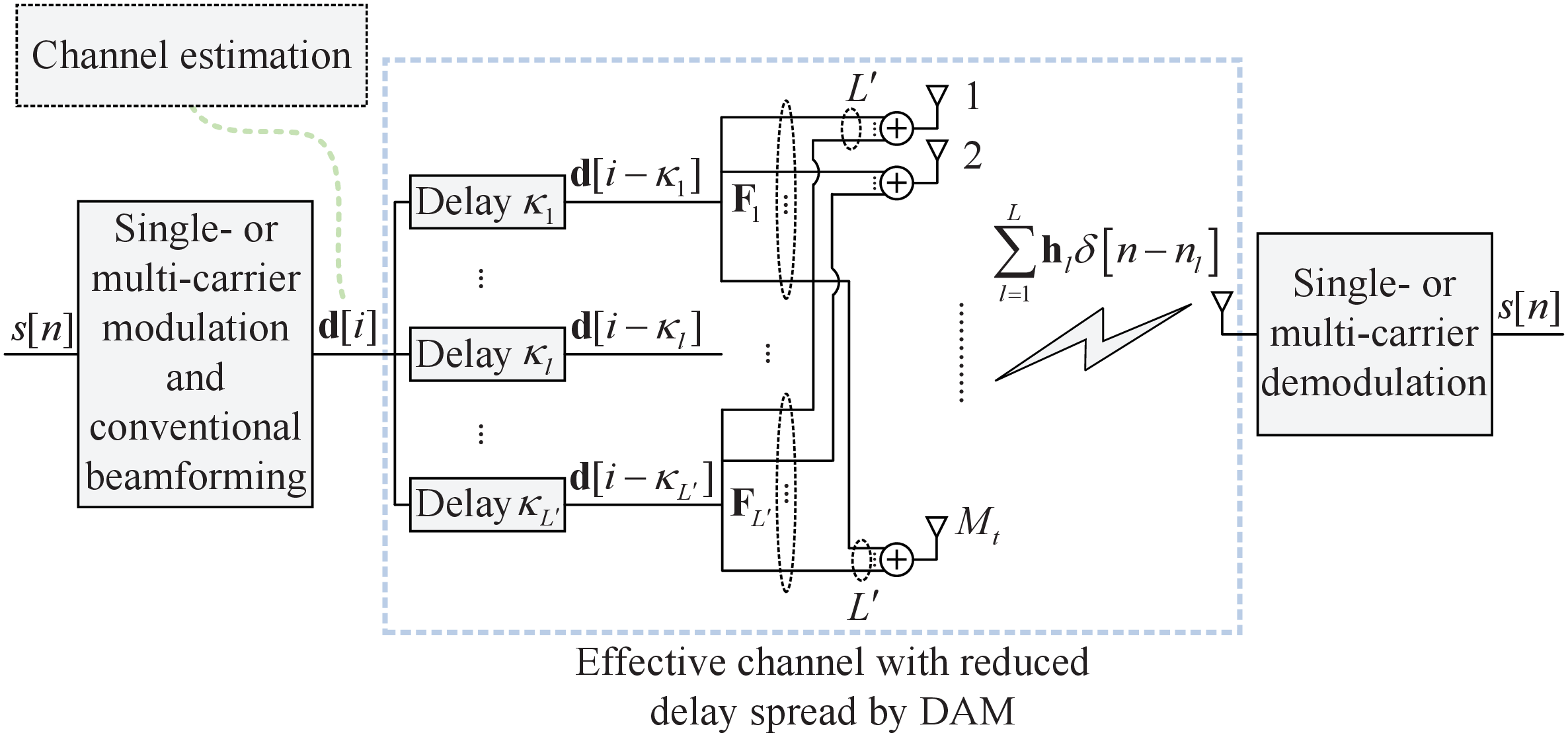}}
 \caption{Single- or multi-carrier communications with DAM processing to reduce the channel delay spread.}
 \label{DAMProcessing}
 \end{figure}

 As shown in Fig~\ref{DAMProcessing}, let ${\bf{d}}\left[ i \right] \in {\mathbb C}^{{M_t} \times 1}$ denote the resulting time-domain signals after either single- or multi-carrier modulation and conventional beamforming, where we use the notation $i$ (rather than $n$ as in the previous section) to denote the time index to account the effect of parallel-to-serial processing in multi-carrier transmission, as will become clear later. For example, the expression of ${\bf{d}}\left[ i \right]$ for DAM-OFDM is given in equation \eqref{newSymbolStreamDAMOFDM} in Section \ref{sectionDAMOFDM}. If the signal ${\bf{d}}\left[ i \right]$ is directly applied to the time-dispersive channel with impulse response given in \eqref{channelImpulseResponse}, the received signal is
 \begin{equation}\label{ReceivedSignalWithoutDAM}
 \begin{aligned}
 y\left[ i \right] &= \sum\limits_{l = 1}^L {{\bf{h}}_l^H{\bf{d}}\left[ {i - {n_l}} \right]}  + z\left[ i \right]\\
 &= \sum\limits_{t = 0}^{{n_{{\rm{span}}}}} {{{\bf{g}}^H}\left[ t \right]{\bf{d}}\left[ {i - t - {n_{\min }}} \right]}  + z\left[ i \right],
 \end{aligned}
 \end{equation}
 where
 \begin{equation}\label{effectiveChannalWithoutDAM}
 {{\bf{g}}^H}\left[ t \right] = \left\{ \begin{split}
 &{\bf{h}}_l^H,\ \ {\rm if}\ \exists l,\ {\rm s.t.}\ t + {n_{\min}} = {n_l},\\
 &{\bf{0}},\ \ \ \ {\rm otherwise}.
 \end{split} \right.
 \end{equation}
 The input-output relationship in \eqref{ReceivedSignalWithoutDAM} clearly shows that the signal ${\bf{d}}\left[ i \right]$ will experience a multi-path channel with delay spread ${n_{\rm span}}={n_{\max}}-{n_{\min}}$. To reduce the effective channel delay spread, we propose the DAM processing technique, as illustrated in Fig.~\ref{DAMProcessing}, for which the time-domain signal transmitted by the $M_t$ antennas is
 \begin{equation}\label{DAMProcessTransmitSignalTime}
 {\bf{\bar q}}\left[ i \right] = \sum\limits_{l = 1}^{L'} {{{{\bf{F}}}_l}{\bf{d}}\left[ {i - {\kappa _l}} \right]},
 \end{equation}
 where ${L'} \le L $ denotes the number of deliberately introduced delay pre-compensations, and ${{\bf{F}}_{l}}\in {{\mathbb {C}}^{M_t \times M_t}}$ denotes the time-domain beamforming matrix associated with the delay pre-compensation $\kappa_l \ge 0$, with ${\kappa _l} \ne {\kappa _{l'}}$, $\forall l \ne l'$. Note that DAM requires CSI to perform path delay pre-compensation and path-based beamforming, as illustrated in Fig~\ref{DAMProcessing}. A preliminary effort on channel estimation for single-carrier DAM is being pursued in our separate work \cite{ding2022channel}, and the efficient channel estimation and feedback for multi-carrier DAM are important to investigate in the future. For the challenging scenario with varying number of channel paths, obtaining the corresponding channel features will be doable with the use of XL-MIMO and the emerging trend of ILSAC, rendering it possible to implement DAM in such a scenario. It is also worth mentioning that the implementation of DAM is on a per-path basis. Benefiting from the double timescales for the time-frequency channel and the parameters of the individual multi-paths (e.g., AoA/AoD and delay), the path delay pre-compensation and path-based beamforming of DAM will remain unchanged within the {\it path invariant time $\bar T$} rather than the channel coherence time $T_c$, with ${\bar T} \gg {T_c}$ \cite{xiao2023exploiting,lu2023delayDoppler}. This thus eases the requirements of channel estimation and the practical implementation of DAM.

 With channel impulse response given in \eqref{channelImpulseResponse}, the received signal with channel input \eqref{DAMProcessTransmitSignalTime} is
 \begin{equation}\label{DAMProcessTimeDomainReceivedSignal}
 \begin{aligned}
 y\left[ i \right] &= \sum\limits_{l = 1}^L {{\bf{h}}_l^H{\bf{\bar q}}\left[ {i - {n_l}} \right]}  + z\left[ i \right]\\
 &= \sum\limits_{l = 1}^L {\sum\limits_{{l'} = 1}^{L'} {{\bf{h}}_l^H{{\bf{{ F}}}_{l'}}{\bf{d}}\left[ {i - {n_l} - {\kappa _{l'}}} \right]}  + z\left[ i \right]}.
 \end{aligned}
 \end{equation}
 The received signal in \eqref{DAMProcessTimeDomainReceivedSignal} is the superposition of $LL'$ components, with delays $\left( {{n_l} + {\kappa _{l'}}} \right)$ and gains ${\bf{h}}_l^H{{\bf{F}}_{l'}}$. It seems that the resulting delay spread will be even increased since more signal components are created. However, the additional design degrees-of-freedom (DoF) of $\{ {{\bf{F}}_{l'}},{\kappa _{l'}}\} _{l' = 1}^{L'}$ make it feasible to actually reduce the resulting multi-path delay spread. The key idea is to design $\{ {\kappa _{l'}}\} _{l' = 1}^{L'}$ to align as many paths as possible to certain desired delay window of spread ${n'_{\rm span}}<{n_{\rm span}}$, and design $\{ {{\bf{F}}_{l'}}\} _{l' = 1}^{L'}$ to zero-force those paths out of the window. Specifically, define the set ${\cal S} \triangleq \left\{ {\left( {l,l'} \right):{\bf{h}}_l^H{{\bf{F}}_{l'}} \ne {\bf{0}}} \right\}$, where $L' \le \left| {\cal S} \right| \le {LL'}$. Then \eqref{DAMProcessTimeDomainReceivedSignal} can be equivalently written as
 \begin{equation}\label{DAMProcessTimeDomainReceivedSignal2}
 y\left[ i \right] = \sum\limits_{\left( {l,l'} \right) \in {\cal S}} {{\bf{h}}_l^H{{\bf{F}}_{l'}}{\bf{d}}\left[ {i  - {n_l} - {\kappa _{l'}}} \right]}  + z\left[ {i} \right].
 \end{equation}
 The resulting delay spread in \eqref{DAMProcessTimeDomainReceivedSignal2} is then given by
 \begin{equation}\label{delaySpreadDAMProcessing}
 {n'_{\rm{span}}} \triangleq \mathop {\max }\limits_{\left( {l,l'} \right) \in {\cal S}} \left( {{n_l} + {\kappa _{l'}}} \right) - \mathop {\min }\limits_{\left( {l,l'} \right) \in {\cal S}} \left( {{n_l} + {\kappa _{l'}}} \right).
 \end{equation}
 By properly designing $\{{{\bf{F}}_{l'}},{\kappa _{l'}}\} _{l' = 1}^{L'}$, DAM aims to reduce the channel delay spread such that ${n'_{\rm{span}}} < {n_{\rm{span}}}$, while still benefiting from the multi-path signal components. Intuitively, for each $l' \in \left[1,L'\right]$, if the time-domain beamforming matrix ${{\bf{F}}_{l'}}$ is designed to zero-force more channels in $\{ {{\bf{h}}_l}\} _{l = 1}^L$, the set ${\cal S}$ will become smaller, which makes it easier to align the resulting delays to achieve a smaller delay spread $n'_{\rm span}$. However, this leads to a more stringent ZF requirement for ${{\bf{F}}_{l'}}$. As such, there exists a trade-off between the desired delay spread and the ZF requirement. In the following, we propose one effective solution to design $\{ {{\bf{F}}_{l'}},{\kappa _{l'}}\} _{l' = 1}^{L'}$ to achieve certain desired channel delay spread $n'_{\rm span} < {n_{\rm span}}$. To gain some insights, we first consider the case to achieve perfect DAM with $n'_{\rm span} = 0$, and then consider the more general case for $0 \le  n'_{\rm span} < n_{\rm span}$.

 \subsection{Perfect DAM}\label{subsectionPerfectDAM}
 Motivated by the examples given in Section \ref{sectionSystemModelAndDAM}, to achieve perfect delay alignment, we let $L'=L$ and ${\kappa _{l'}} = {n_{\max }} - {n_{l'}}$, $l' \in \left[ {1,L'} \right]$. Furthermore, the beamforming matrices $\left\{ {{{\bf{F}}_{l'}}} \right\}_{l' = 1}^{L'}$ are designed such that
 \begin{equation}\label{DAMProcessZFCondition}
 {{\bf{h}}_l^H}{{\bf{F}}_{l'}} = {{\bf{0}}_{1 \times M_t}},\ \forall l \ne l'.
 \end{equation}
 Similar to \eqref{DAMCondition}, the ZF conditions in \eqref{DAMProcessZFCondition} can be compactly written as ${\bf{H}}_{l'}^H{{\bf{F}}_{l'}} = {{\bf{0}}_{\left( {L - 1} \right) \times {M_t}}}$, $\forall l'$, with ${{\bf{H}}_{l'}}$ defined below \eqref{DAMCondition}, which are feasible almost surely for $M_t \ge L$. The beamforming matrix ${{\bf{F}}_{l'}}$ is then expressed as ${{\bf{F}}_{l'}} = {\bf{H}}_{l'}^ \bot {{\bf{X}}_{l'}}$, with ${{\bf{X}}_{l'}} \in {{\mathbb C}^{{r_{l'}} \times {M_t}}}$. In this case, the path set ${\cal S}$ in \eqref{delaySpreadDAMProcessing} is given by ${\cal S} = \left\{ {\left( {l,l} \right):1 \le l \le L} \right\}$, and it is not difficult to see from \eqref{delaySpreadDAMProcessing} that $n'_{\rm span}=0$. Furthermore, the received signal in \eqref{DAMProcessTimeDomainReceivedSignal2} reduces to
 \begin{equation}\label{DAMProcessTimeDomainReceivedSignalPerfect}
 y\left[ i \right] = \left( {\sum\limits_{l = 1}^L {{\bf{h}}_l^H{\bf{H}}_l^ \bot {\bf{X}}_l} } \right){\bf{d}}\left[ {i - {n_{\max }}} \right] + z\left[ i \right].
 \end{equation}
 It is observed from \eqref{DAMProcessTimeDomainReceivedSignalPerfect} that similar to \eqref{reducedDelayCompensationReceivedSignalMRT} and \eqref{DelayCompensationReceivedSignalEquivalent}, the original time-dispersive channel with delay spread $n_{\rm span}$ has been transformed to the non-dispersive channel, which eases the single- or multi-carrier modulation. For instance, it is not difficult to see that by letting ${\bf{d}}\left[ i \right] = {\bf{v}}s\left[ i \right]$, where ${\bf{v}} \in {{\mathbb C}^{{M_t} \times 1}}$, we have the ISI-free single-carrier system equivalent to \eqref{DelayCompensationReceivedSignalEquivalent}.

 \subsection{Generic DAM}\label{subsectionGeneralDAM}
\begin{figure}[!t]
 \centering
 \centerline{\includegraphics[width=3.5in,height=2.15in]{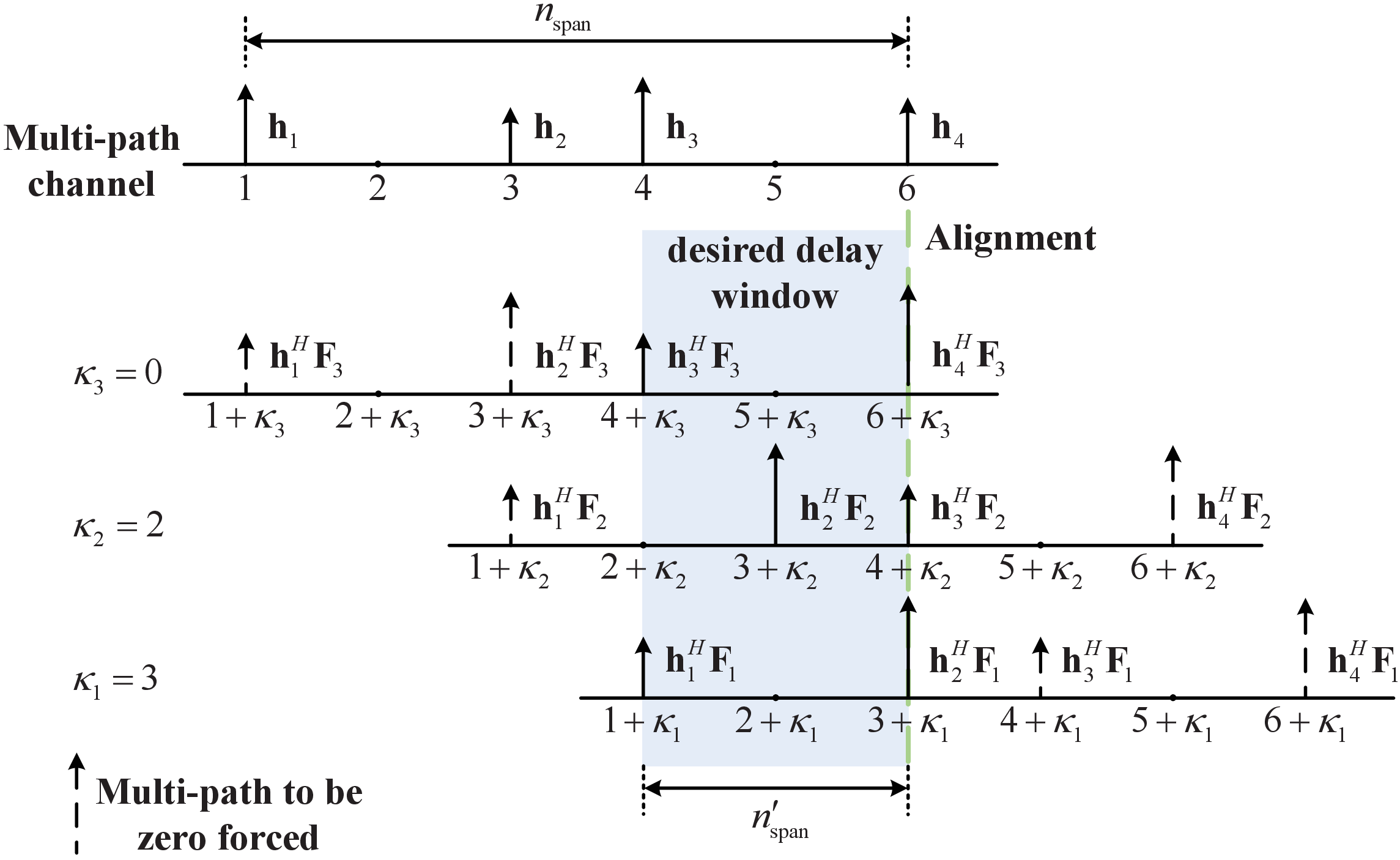}}
 \caption{An illustration of DAM to reduce the channel delay spread. For the original channel, the number of multi-paths is $L=4$, with ${n_1} = 1$, ${n_2} = 3$, ${n_3} = 4$, and ${n_4} = 6$, and the delay spread is $n_{\rm span} = 5$. The number of deliberately introduced delay pre-compensations is $L' =3$, which are designed such that ${n_4} + {\kappa _3} = {n_3} + {\kappa _2} = {n_2} + {\kappa _1} = 6$, i.e., $\kappa _3 = 0$, $\kappa _2 = 2$, and $\kappa _1 = 3$. All components outside of the desired delay window are forced to zero by ensuring ${\bf{\bar H}}_{l'}^H{{{\bf{ F}}}_{l'}} = {{\bf{0}}}$. The resulting delay spread is reduced to $n'_{\rm span}=2$.}
 \label{reducedDelaySpreadIllustration}
 \end{figure}

 When perfect DAM is infeasible or undesirable, the more generic DAM scheme can be applied to achieve certain desired delay spread ${n'_{\rm span}}<{n_{\rm span}}$. To this end, instead of letting ${n_{l'}} + {\kappa _{l'}} = {n_{\max }}$ as in Section \ref{subsectionPerfectDAM}, the delay pre-compensations $\{ {\kappa _{l'}}\} _{l' = 1}^{L'}$ are designed so that ${n_{L - L' + l'}} + {\kappa _{l'}} = {n_{\max }}$, $\forall l' \in \left[1,L'\right]$. A simple illustration is given in Fig.~\ref{reducedDelaySpreadIllustration}, with $L = 4$ multi-paths and $L'=3$ deliberately introduced delay pre-compensations.
 For each introduced delay pre-compensation $\kappa_{l'}$, all the $L$ multi-path delays are shifted by the common value $\kappa_{l'}$. The green dotted line shows the aligned delay terms, and the blue box encloses the desired delay window, which includes all the $L=4$ multi-path components and has width equal to the desired delay spread ${n'_{\rm span}}<{n_{\rm span}}$. Based on the chosen window, the transmit beamforming matrices $\{ {{{\bf{F}}_{l'}}} \}_{l' = 1}^{L'}$ are designed so that all components outside the desired delay window are forced to zero. To this end, for any $l' \in \left[ 1,L'\right]$, the multi-paths outside and inside the desired delay window are respectively given by
 \begin{equation}\label{ineffectiveMultiPath}
 {{\cal L}_{l'}} \triangleq \left\{ {1 \le l \le L:{n_l} + {\kappa _{l'}} \notin \left[ {{n_{\max }} - {n'_{{\rm{span}}}},{n_{\max }}} \right]} \right\},
 \end{equation}
 \begin{equation}\label{effectiveMultiPath}
 {\bar {\cal L}}_{l'} \triangleq \left\{ {1 \le l \le L:{n_l} + {\kappa _{l'}} \in \left[ {{n_{\max }} - {n'_{{\rm{span}}}},{n_{\max }}} \right]} \right\},
 \end{equation}
 where $L - 1 - n'_{\rm{span}} \le \left| {{\cal L}_{l'}} \right| \le L - 1$, and ${\bar {\cal L}}_{l'}$ is the complementary set of ${{\cal L}_{l'}}$,  i.e., ${{\cal L}_{l'}} \cup {\bar {\cal L}}_{l'} = \{l: 1\le l \le L\}$. For each $1\le l' \le L'$, define ${{\bf{\bar H}}_{l'}} \in {{\mathbb C}^{{M_t} \times \left| {{\cal L}_{l'}} \right|}} \triangleq \left[ {{{\bf{h}}_l}} \right]_{l \in {{\cal L}_{l'}}} $. To design ${\bf{F}}_{l'}$ so that those multi-paths outside the desired delay window are eliminated, we should have
 \begin{equation}\label{ZFConditionDAMProcessing}
  {\bf{\bar H}}_{l'}^H{{{\bf{ F}}}_{l'}} = {{\bf{0}}_{\left| {{\cal L}_{l'}} \right| \times {M_t}}}.
 \end{equation}
 The above condition is feasible almost surely as long as ${M_t} \ge {\left| {{\cal L}_{l'}} \right|} + 1$, which can always be achieved by choosing the appropriate delay spread $n'_{\rm span}$ even when ${M_t}<L$. Similar to \eqref{DAMCondition} and \eqref{DAMProcessZFCondition}, the beamforming matrix is expressed as ${{\bf{F}}_{l'}} = {\bf{\bar H}}_{l'}^ \bot {\bf{\bar X}}_{l'}$, where ${\bf{\bar H}}_{l'}^ \bot  \in {{\mathbb C}^{{M_t} \times {{\bar r}_{l'}}}}$ is an orthonormal basis for the orthogonal complement of ${{{\bf{\bar H}}}_{l'}}$, and ${{{\bf{\bar X}}}_{l'}} \in {{\mathbb C}^{{{\bar r}_{l'}} \times {M_t} }}$ is the new beamforming matrix to be designed, with ${{\bar r}_{l'}} = {\rm{rank}}( {{\bf{\bar H}}_{l'}^ \bot } ) = {M_t} - \left| {{\cal L}_{l'}} \right|$.

 As a result, the received signal in \eqref{DAMProcessTimeDomainReceivedSignal2} can be expressed as
 \begin{equation}\label{DAMProcessTimeDomainReceivedSignalGeneral1}
 y\left[ i \right] = \sum\limits_{l' = 1}^{L'} {\sum\limits_{l \in {{\bar {\cal L}}_{l'}}} {{\bf{h}}_l^H{\bf{\bar H}}_{l'}^ \bot {\bf{\bar X}}_{l'}{\bf{d}}\left[ {i - {n_l} - {\kappa _{l'}}} \right]} }  + z\left[ i \right],
 \end{equation}
 where ${\kappa _{l'}} = {n_{\max }} - {n_{L - L' + l'}}$. To more clearly show the impact of the resulting delay spread, we should group those multi-paths with common delays \cite{zeng2018multi,lu2022delay}. To this end, for any $t \in \left[ {0,{n'_{\rm{span}}}} \right]$, define the following effective channel vector
 \begin{equation}\label{newDefinitionEffectiveChannelVector}
 {\bf{g}}_{l'}^H\left[ t \right] {\rm =} \left\{ \begin{split}
 &{\bf{h}}_l^H,\ {\rm if}\ \exists l \in {{\bar {\cal L}}_{l'}},\ {\rm s.t.}\ {n_l} + {\kappa _{l'}} = t + {{n_{\max }} - {n'_{\rm{span}}}},\\
 &{\bf{0}},\ \ \ {\rm{otherwise}}.
 \end{split} \right.
 \end{equation}
 For example, for $l'=1$ and $t=0$ in Fig.~\ref{reducedDelaySpreadIllustration}, since ${n_1} + {\kappa _1} = 0 + {n_{\max }} - {n'_{\rm{span}}}$, we have ${\bf{g}}_1^H\left[ 0 \right] = {\bf{h}}_1^H$. Similarly, ${\bf{g}}_1^H[ {{n'_{\rm{span}}}} ] = {\bf{h}}_2^H$. Then \eqref{DAMProcessTimeDomainReceivedSignalGeneral1} can be equivalently written as
 \begin{equation}\label{DAMProcessTimeDomainReceivedSignalGeneral2}
 \begin{aligned}
 y\left[ i \right]  & {\rm =} \sum\limits_{t = 0}^{{n'_{{\rm{span}}}}} {\left( {\sum\limits_{l' = 1}^{L'} {{\bf{g}}_{l'}^H\left[ t \right]{{\bf{\bar H}}_{l'}^ \bot {\bf{\bar X}}_{l'} }} } \right){{\bf{d}}\left[ {i  - t  - ( {{n_{\max }} - {n'_{\rm{span}}}} )} \right]}}\\
 &\ \ \ \ \ \ \ \ \ \ \ \ \ \ \ \ \ \ \ \ \ \ \ \ \ \ \ \ \ \ \ \ \ \ \ \ \ \ \ \ \ \ \ \ \ \ \ \  + z\left[ {i} \right].
 \end{aligned}
 \end{equation}
 Since ${n_{\max }} - {n'_{{\rm{span}}}}$ is the minimum delay that is common to all multi-paths, by simple index substitution, \eqref{DAMProcessTimeDomainReceivedSignalGeneral2} can be equivalently expressed as
 \begin{equation}\label{DAMProcessTimeDomainReceivedSignalGeneral3}
  y\left[ i \right] = \sum\limits_{t = 0}^{{n'_{{\rm{span}}}}} {\left( {\sum\limits_{l' = 1}^{L'} {{\bf{g}}_{l'}^H\left[ t \right]{{\bf{\bar H}}_{l'}^ \bot {\bf{\bar X}}_{l'} }} } \right){{\bf{d}}\left[ {i - t } \right]}}+ z\left[ {i} \right].
 \end{equation}
 It is observed from \eqref{ReceivedSignalWithoutDAM} and \eqref{DAMProcessTimeDomainReceivedSignalGeneral3} that after DAM processing, the single- or multi-carrier signal ${{\bf{d}}\left[ i \right]}$ would see an effective channel whose channel delay spread reduced from $n_{\rm span}$ to our selected value $n'_{\rm{span}}$, while still benefiting from the multi-path signal components. As a result, the proposed DAM achieves the ISI mitigation while exploiting the multi-path signal components. This thus provides a new DoF to combat channel time dispersion to enable more efficient single- or multi-carrier transmissions, as will be illustrated by the proposed DAM-OFDM technique in the next section.

\section{DAM Meets OFDM}\label{sectionDAMOFDM}
 In this section, to illustrate the benefits brought by DAM, we propose the novel DAM-OFDM technique, which integrates DAM to flexibly manipulate the channel delay spread for more efficient OFDM transmission. Specifically, with reduced channel delay spread, the DAM-OFDM can save the CP overhead for the same number of sub-carriers, or mitigate the practical issue of PAPR by using fewer sub-carriers, yet without increasing the CP overhead.

 \begin{figure*}[!t]
 \centering
 \centerline{\includegraphics[width=6.3in,height=2.9in]{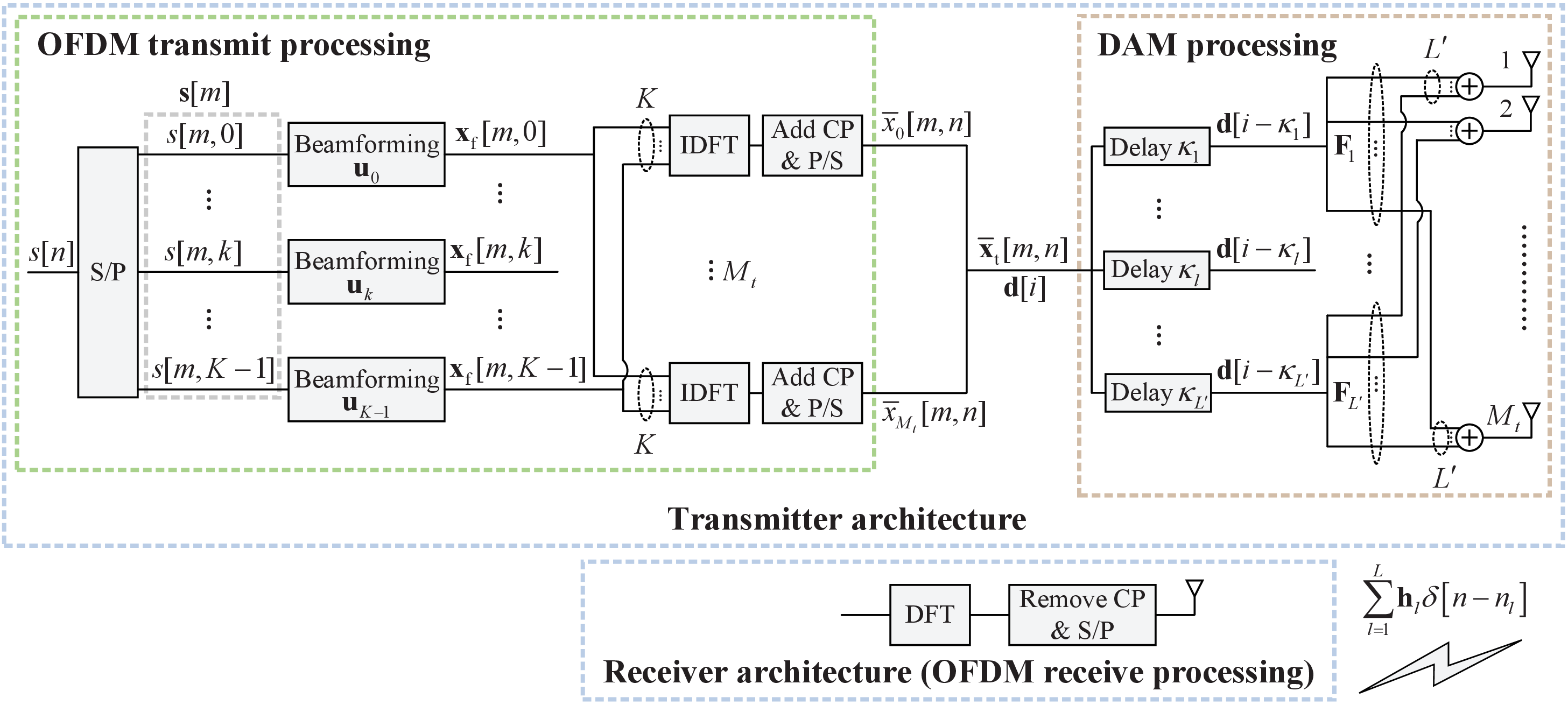}}
 \caption{Transceiver architecture of the proposed DAM-OFDM communication system.}
 \label{MISOOFDMDAM}
 \end{figure*}
 As illustrated in Fig.~\ref{MISOOFDMDAM}, let ${\bf{s}}\left[ m \right]  \in {{\mathbb {C}}^{{K} \times 1}}  \triangleq \left[ {s\left[ {m,0} \right],} \right.$ ${\left. { \cdots ,s\left[ {m,K - 1} \right]} \right]^T}$ denote the information-bearing symbols of the $m$th OFDM symbol, where $K$ is the number of OFDM sub-carriers, and $s\left[ {m,k} \right]$ is obtained from the information-bearing symbols $s\left[ n \right]$ in Section \ref{sectionSystemModelAndDAM} that are arranged in time-frequency domain for the $k$th sub-carrier of the $m$th OFDM symbol, with ${\mathbb E}[ {{{\left| {s[m,k]} \right|}^2}} ] = 1,\ 0 \le k \le K - 1$. Denote by ${{\bf{u}}_k} \in {{\mathbb {C}}^{M_t \times 1}}$ the frequency-domain beamforming vector for the $k$th sub-carrier. Then the signal in the frequency domain is ${{\bf{x}}_{\rm f}}\left[ {m,k} \right] = {{\bf{u}}_k}s\left[ {m,k} \right]$. Let ${{\bf{x}}_{\rm t}}\left[ {m,n} \right]\in {{\mathbb {C}}^{M_t \times 1}}$ be the $n$th sample of the $m$th OFDM symbol in the time-domain, which is obtained by applying $K$-point inverse discrete Fourier transform (IDFT) to ${{\bf{x}}_{\rm f}}\left[ {m,k} \right]$, given by
 \begin{equation}\label{timeDomainTransmitSignal}
 \begin{aligned}
 {{\bf{x}}_{\rm t}}\left[ {m,n} \right] &= \frac{1}{{\sqrt K }}\sum\limits_{k = 0}^{K - 1} {{{\bf{x}}_{\rm f}}\left[ {m,k} \right]{e^{j\frac{{2\pi }}{K}kn}}}\\
 &= \frac{1}{{\sqrt K }}\sum\limits_{k = 0}^{K - 1} {{{\bf{u}}_k}s\left[ {m,k} \right]{e^{j\frac{{2\pi }}{K}kn}}},\ n = 0, \cdots ,K - 1.
 \end{aligned}
 \end{equation}
 The time-domain signal after CP insertion is
 \begin{equation}\label{timeDomainTransmitSignalWithCP}
 {{\bf{\bar x}}_{\rm t}}\left[m, n \right] = \left\{ \begin{split}
 &{{\bf{x}}_{\rm t}}\left[m, n \right],\ \ \ \ \ \ \ \ n = 0, \cdots ,K - 1,\\
 &{{\bf{x}}_{\rm t}}\left[m, {n + K} \right],\ \ n =  - {{N}_{\rm CP}}, \cdots , - 1.
 \end{split} \right.
 \end{equation}
 As discussed in Section \ref{sectionManipulatingDelaySpread},  with DAM processing, the channel delay spread has been reduced from $n_{\rm span}$ to $n'_{\rm span}$. Therefore, the length of the CP for DAM-OFDM only needs to satisfy ${n'_{\rm span}} \le {N_{\rm{CP}}} \le K$. By substituting \eqref{timeDomainTransmitSignal} into \eqref{timeDomainTransmitSignalWithCP} and using the identity ${e^{j\frac{{2\pi }}{K}k\left( {n + K} \right)}} = {e^{j\frac{{2\pi }}{K}kn}}$, ${{\bf{\bar x}}}_{\rm t}\left[m, n \right]$ can be compactly written as
 \begin{equation}\label{timeDomainTransmitSignalWithCPSpecific}
 {{\bf{\bar x}}_{\rm t}}\left[m, n \right] {\rm =} \frac{1}{\sqrt K }\sum\limits_{k = 0}^{K-1} {{{\bf{u}}_k}s\left[m, k \right]{e^{j\frac{{2\pi }}{K}kn}}},\ n {\rm =}  - {{N}_{\rm CP }}. \cdots ,K - 1.
 \end{equation}
 With parallel to serial conversion, the samples of all OFDM symbols ${{\bf{\bar x}}_{\rm t}}\left[m, n \right]$ are concatenated as a time-domain sequence ${\bf{d}}\left[ i \right]$, as defined in Section \ref{sectionManipulatingDelaySpread}, which is given by
 \begin{equation}\label{newSymbolStreamDAMOFDM}
 {\bf{d}}\left[ i \right] \triangleq {{{\bf{\bar x}}}_{\rm{t}}}\left[ {m,n} \right], \ \forall n \in \left[ { - {N_{\rm{CP}}},K - 1} \right],
 \end{equation}
 where $i = m\left( {K + {N_{{\rm{CP}}}}} \right) + n$. On the other hand, with any given time index $i$, its corresponding OFDM symbol and sample number can be respectively obtained as
 \begin{equation}\label{iCorrespondingm}
 m = \left\lfloor {\frac{{i + {N_{{\rm{CP}}}}}}{{K + {N_{{\rm{CP}}}}}}} \right\rfloor,
 \end{equation}
 \begin{equation}\label{iCorrespondingn}
 n = {\rm mod} \left( {i + {N_{{\rm{CP}}}},K + {N_{{\rm{CP}}}}} \right) - {N_{{\rm{CP}}}}.
 \end{equation}

 With DAM-OFDM, the time-domain transmitted signal ${\bf{\bar q}}\left[ i \right]$ in \eqref{DAMProcessTransmitSignalTime} is obtained by substituting ${\bf{d}}\left[ i \right]$ with \eqref{newSymbolStreamDAMOFDM}, which is given by
 \begin{equation}\label{DAMOFDMTransmitSignalTime}
 {\bf{\bar q}}\left[ i \right] = \sum\limits_{l = 1}^{L'} {{{\bf{\bar H}}_{l}^ \bot {{{\bf{\bar X}}}_{l}}}{\bf{d}}\left[ {i - {\kappa _l}} \right]},
 \end{equation}
 where we have used ${{\bf{F}}_{l}} = {\bf{\bar H}}_{l}^  \bot {{{\bf{\bar X}}}_{l}}$ as presented below \eqref{ZFConditionDAMProcessing}. Based on \eqref{iCorrespondingm} and \eqref{iCorrespondingn}, ${\bf{d}}\left[ {i - {\kappa _l}} \right]$ in terms of OFDM symbol and sample number can be expressed as
 \begin{equation}\label{diMinustaosymbolStreamDAMOFDMSymSamNum}
 \begin{aligned}
 {\bf{d}}\left[ {i - {\kappa _l} } \right]& =  {{{\bf{\bar x}}}_{\rm{t}}}\Big[ {\left\lfloor {\frac{{i - {\kappa _l} + {N_{\rm{CP}}}}}{{K + {N_{\rm{CP}}}}}} \right\rfloor ,} \Big.\\
 &\ \ \ \ \Big. {{\rm{mod}}\left( {i -{\kappa _l}  + {N_{\rm{CP}}},K + {N_{\rm{CP}}}} \right) - {N_{\rm{CP}}}} \Big].
 \end{aligned}
 \end{equation}

 \begin{theorem}\label{TheoremTransmitPowerDAMOFDMSignal}
 The transmit power of ${\bf{\bar q}}\left[i \right]$ in \eqref{DAMOFDMTransmitSignalTime} is
 \begin{equation}\label{transmitPowerDAMOFDMSignal}
 {\mathbb E}\left[ {{{\left\| {{\bf{\bar q}}\left[i \right]} \right\|}^2}} \right] = \frac{1}{K}\sum\limits_{k = 0}^{K - 1} {{{\Big\| {\sum\limits_{{l} = 1}^{L'} {{{\bf{\bar H}}_{l}^ \bot {\bf{\bar X}}_{l} }{{\bf{u}}_k}{e^{ - j\frac{{2\pi }}{{K}}k{\kappa _{l}}}}} } \Big\|^2}}}.
 \end{equation}
 \end{theorem}
 \begin{IEEEproof}
 Please refer to Appendix~\ref{proofOfTheoremTransmitPowerDAMOFDMSignal}.
 \end{IEEEproof}

 Furthermore, by substituting $i = m\left( {K + {N_{\rm{CP}}}} \right) + n$ into the obtained input-output relationship in \eqref{DAMProcessTimeDomainReceivedSignalGeneral3} of Section \ref{subsectionGeneralDAM}, the $\left(K+{N_{\rm CP}}\right)$ samples of the $m$th OFDM symbol, including the CP, are given by \eqref{nthReceivedSampleOfmthsymbolDAMOFDM}, shown at the top of this page.
 \newcounter{mytempeqncnt1}
 \begin{figure*}
 \normalsize
 \setcounter{mytempeqncnt1}{\value{equation}}
 \begin{equation}\label{nthReceivedSampleOfmthsymbolDAMOFDM}
 \begin{aligned}
 y\left[ {m,n} \right] = & \sum\limits_{t = 0}^{{n'_{{\rm{span}}}}} {\left( {\sum\limits_{l' = 1}^{L'} {{\bf{g}}_{l'}^H\left[ t \right]{{\bf{\bar H}}_{l'}^ \bot {\bf{\bar X}}_{l'} }    } } \right){{{\bf{\bar x}}}_t}\Big[ {\left\lfloor {\frac{{m\left( {K + {N_{{\rm{CP}}}}} \right) + n - t + {N_{{\rm{CP}}}}}}{{K + {N_{{\rm{CP}}}}}}} \right\rfloor }, \Big.} \\
 &\ \ \ \ \ \ \ \ \ \ \ \ \ \ \ \Big. {{\rm{mod}}\left( {m\left( {K + {N_{{\rm{CP}}}}} \right) + n - t + {N_{{\rm{CP}}}},K + {N_{{\rm{CP}}}}} \right) - {N_{{\rm{CP}}}}} \Big] + z\left[ {m,n}  \right], \ \forall n \in \left[ { - {N_{\rm{CP}}},K - 1} \right].
 \end{aligned}
 \end{equation}
 \hrulefill
 \vspace{-3ex}
 \end{figure*}
 With ${N_{\rm CP}} \ge {n'_{\rm span}}$, it follows from \eqref{nthReceivedSampleOfmthsymbolDAMOFDM} that for the last $K$ samples $n \in \left[0,K-1\right]$, the reception of the $m$th OFDM symbol $y\left[ {m,n} \right]$ is free from the inter-block interference of the $\left(m-1\right)$th OFDM symbol. To see this, by discarding the CP that corresponds to the first $N_{\rm CP}$ samples in \eqref{nthReceivedSampleOfmthsymbolDAMOFDM}, we have
 \begin{equation}
 \begin{aligned}
 &\left\lfloor {\frac{{m\left( {K + {N_{{\rm{CP}}}}} \right) + n - t + {N_{{\rm{CP}}}}}}{{K + {N_{{\rm{CP}}}}}}} \right\rfloor  = m,\\
 &{\rm{mod}}\left( {m\left( {K + {N_{{\rm{CP}}}}} \right) + n - t + {N_{{\rm{CP}}}},K + {N_{{\rm{CP}}}}} \right)\\
 &\ \ \ \ \ \ \ \ \ \ \ \ \ \ - {N_{{\rm{CP}}}} = n-t, \ \forall n \in \left[ {0,K - 1} \right].
 \end{aligned}
 \end{equation}
 As a result, for the proposed DAM-OFDM with ${N_{\rm{CP}}} \ge {n'_{\rm span}}$, the time-domain received signal in \eqref{nthReceivedSampleOfmthsymbolDAMOFDM} after removing the CP can be expressed as \eqref{DAMOFDMReceiveSignalTime}, shown at the top of the next page,
 \newcounter{mytempeqncnt2}
 \begin{figure*}
 \normalsize
 \setcounter{mytempeqncnt2}{\value{equation}}
 \begin{equation}\label{DAMOFDMReceiveSignalTime}
 \begin{aligned}
 y\left[ {m,n} \right] & = \sum\limits_{t = 0}^{{n'_{{\rm{span}}}}} {\left( {\sum\limits_{l' = 1}^{L'} {{\bf{g}}_{l'}^H\left[ t \right]{{\bf{\bar H}}_{l'}^ \bot {\bf{\bar X}}_{l'}}}} \right){{{\bf{\bar x}}}_t}\left[ {m,n - t} \right]}+ z\left[ {m,n}  \right]\\
 &= \sum\limits_{t = 0}^{{n'_{{\rm{span}}}}} {\left( {\sum\limits_{l' = 1}^{L'} {{\bf{g}}_{l'}^H\left[ t \right]{{\bf{\bar H}}_{l'}^ \bot {\bf{\bar X}}_{l'} }} } \right)} \frac{1}{{\sqrt {K} }}\sum\limits_{k = 0}^{K - 1} {{{\bf{u}}_k}s\left[ {m,k} \right]{e^{j\frac{{2\pi }}{{K}}k\left( {n - t} \right)}}}+ z\left[ {m,n}  \right] \\
 & = \frac{1}{{\sqrt {K} }}\sum\limits_{k = 0}^{K - 1} {\underbrace {\left( {\sum\limits_{t = 0}^{{n'_{{\rm{span}}}}} {\left( {\sum\limits_{l' = 1}^{L'} {{\bf{g}}_{l'}^H\left[ t \right]{{\bf{\bar H}}_{l'}^ \bot {\bf{\bar X}}_{l'} }} } \right){e^{ - j\frac{{2\pi }}{{K}}kt}}} } \right)}_{ \sqrt K {{\bf{\tilde h}}^H}\left[ k \right]  }{{\bf{u}}_k}s\left[ {m,k} \right]{e^{j\frac{{2\pi }}{{K}}kn}}}+ z\left[ {m,n}  \right],\ \forall n \in \left[ {0,K - 1} \right].
 \end{aligned}
 \end{equation}
 \hrulefill
 \end{figure*}
 where the second equality is obtained by substituting ${{\bf{\bar x}}_{\rm t}}\left[m, n \right]$ with \eqref{timeDomainTransmitSignalWithCPSpecific}, and ${{{\bf{\tilde h}}}^H}\left[ k \right]$ denotes the frequency-domain channel of the $k$th sub-carrier for DAM-OFDM, which is given by
 \begin{equation}\label{effectiveFrequencyDomainChannelDAMOFDM}
 {{{\bf{\tilde h}}}^H}\left[ k \right] = \frac{1}{{\sqrt K }}\sum\limits_{t = 0}^{{n'_{{\rm{span}}}}} {\left( {\sum\limits_{l' = 1}^{L'} {{\bf{g}}_{l'}^H\left[ t \right]{\bf{\bar H}}_{l'}^ \bot {{{\bf{\bar X}}}_{l'}}} } \right){e^{ - j\frac{{2\pi }}{K}kt}}} , \ \forall k,
 \end{equation}
 where the coefficient $1/{\sqrt K}$ is introduced to ensure the conservation of energy between DFT and IDFT. In particular, for the conventional OFDM without DAM, we have $L'=1$, ${\kappa _{1}} = 0$, ${\bf{\bar H}}_1^ \bot {{{\bf{\bar X}}}_1} = {\bf{I}}$, and $n'_{\rm span} = n_{\rm span}$. According to the definition in \eqref{newDefinitionEffectiveChannelVector}, the frequency-domain channel reduces to ${{{\bf{\tilde h}}}^H}\left[ k \right] = \frac{1}{{\sqrt K }}\sum\nolimits_{t = 0}^{{n_{{\rm{span}}}}} {{\bf{g}}_1^H\left[ t \right]{e^{ - j\frac{{2\pi }}{K}kt}}}  = \frac{1}{{\sqrt K }}\sum\nolimits_{l = 1}^L {{\bf{h}}_l^H} {e^{j\frac{{2\pi }}{K}k\left( {{n_{\min }} - {n_l}} \right)}}$. Therefore, the proposed DAM-OFDM generalizes the conventional OFDM, since it includes the conventional OFDM as a special case.

 The received signal in the frequency-domain can be obtained by taking the DFT to \eqref{DAMOFDMReceiveSignalTime}, which yields
 \begin{equation}\label{DAMOFDMReceiveSignalFrequency}
 {y_{\rm f}}\left[ {m,k} \right] = {{{\bf{\tilde h}}}^H}\left[ k \right]{{\bf{u}}_k}s\left[ {m,k} \right] + z\left[m,k\right], \ \forall k \in \left[0, {K}-1\right].
 \end{equation}
 Therefore, the received SNR for sub-carrier $k$ of DAM-OFDM can be expressed as
 \begin{equation}\label{generalCaseSubCarrierkSNR}
 {\gamma _k} = \frac{{{{\left| {{{{\bf{\tilde h}}}^H}\left[ k \right]{{\bf{u}}_k}} \right|}^2}}}{{{\sigma }^2}/K} =\frac{{{{\Big| {\sum\limits_{t = 0}^{{n'_{{\rm{span}}}}} {\Big( {\sum\limits_{l' = 1}^{L'} {{\bf{g}}_{l'}^H\left[ t \right]{{\bf{\bar H}}_{l'}^ \bot {\bf{\bar X}}_{l'}} } } \Big){e^{ - j\frac{{2\pi }}{{K}}kt}}} {{\bf{u}}_k}} \Big|^2}}}}{{\sigma }^2}.
 \end{equation}

 As a result, the spectral efficiency of DAM-OFDM can be maximized by jointly optimizing the time-domain beamforming matrices $\{ {{{\bf{\bar X}}}_{l'}}\} _{l' = 1}^{L'}$ and the frequency-domain beamforming vectors $\{ {{\bf{u}}_k}\} _{k = 0}^{K - 1}$. The optimization problem can be formulated as
 \begin{equation}
 \begin{aligned}
 \left({\rm P1}\right)& \ \mathop {\max }\limits_{\{ {{\bf{\bar X}}_{l'}}\} _{l' = 1}^{L'},\{ {{\bf{u}}_k}\} _{k = 0}^{K - 1}} \frac{1}{K}\sum\limits_{k = 0}^{K - 1} {{{\log }_2}\left( {1 + {\gamma _k}} \right)} \\
 &\ \ \ \ \ {\rm{s.t.}}\ \sum\limits_{k = 0}^{K - 1} {{{\Big\| {\sum\limits_{l' = 1}^{L'} {{{\bf{\bar H}}_{l'}^ \bot {\bf{\bar X}}_{l'}}{{\bf{u}}_k}{e^{ - j\frac{{2\pi }}{K}k{\kappa _{l'}}}}} } \Big\|^2}}}  \le KP,
 \nonumber
 \end{aligned}
 \end{equation}
 where the power constraint follows from Theorem \ref{TheoremTransmitPowerDAMOFDMSignal}. Problem (P1) is difficult to be directly solved since it is non-convex. Besides, the time-domain beamforming matrices $\{ {{{\bf{\bar X}}}_{l'}}\} _{l' = 1}^{L'}$ and the frequency-domain beamforming vectors $\{{{\bf{u}}_k}\} _{k = 0}^{K - 1}$ are intricately coupled with each other both in the objective function and the power constraint. To solve problem (P1), the SNR in \eqref{generalCaseSubCarrierkSNR} for sub-carrier $k$ is compactly rewritten as
 \begin{equation}\label{frequencyDomainSNRDAMOFDM}
 \begin{aligned}
 {\gamma _k} &= \frac{{{{\Big| {\sum\limits_{l' = 1}^{L'} {\Big( {\sum\limits_{t = 0}^{{n'_{\rm{span}}}} {{\bf{g}}_{l'}^H\left[ t \right]{e^{ - j\frac{{2\pi }}{{K}}kt}}} } \Big){\bf{\bar H}}_{l'}^ \bot {\bf{\bar X}}_{l'}} {{\bf{u}}_k}} \Big|^2}}}}{{\sigma }^2}\\
 &= {\Big| {\sum\limits_{l' = 1}^{L'} {{\bf{g}}_{kl'}^H{\bf{\bar H}}_{l'}^ \bot {{{\bf{\bar X}}}_{l'}}} {{\bf{u}}_k}} \Big|^2} = {\left| {{\bf{g}}_k^H{\bf{\bar X}}{{\bf{u}}_k}} \right|^2} = {\left| {{\bf{g}}_k^H{{\bf{w}}_k}} \right|^2},
 \end{aligned}
 \end{equation}
 where ${\bf{g}}_{kl'} \in {\mathbb C}^{{M_t} \times 1} \triangleq (\sum\nolimits_{t = 0}^{{n'_{\rm{span}}}} {{\bf{g}}_{l'}\left[ t \right]{e^{ j\frac{{2\pi }}{{K}}kt}}})/{\sigma }$, ${\bf{g}}_k \in {\mathbb C}^{({\sum {{\bar r}_{l'}}}) \times 1 }$ $\triangleq [{\bf{g}}_{k1}^H{\bf{\bar H}}_1^ \bot , \cdots ,{\bf{g}}_{kL'}^H{\bf{\bar H}}_{L'}^ \bot ]^H$, ${\bf{\bar X}} \in {{\mathbb C}^{(\sum {{\bar r}_{l'}}) \times {M_t} }} \triangleq {[{\bf{\bar X}}_1^T, \cdots ,{\bf{\bar X}}_{L'}^T]^T}$, and ${{\bf{w}}_k} \in {\mathbb C}^{({\sum {{\bar r}_{l'}}}) \times 1 } \triangleq {{\bf{\bar X}}}{{\bf{u}}_k}$. Besides, the left hand side of the power constraint in problem (P1)
 \begin{equation}
 \begin{aligned}
 &\sum\limits_{k = 0}^{K - 1} {{{\Big\| {\sum\limits_{l' = 1}^{L'} {{\bf{\bar H}}_{l'}^ \bot {\bf{\bar X}}_{l'}{{\bf{u}}_k}{e^{ - j\frac{{2\pi }}{{K}}k{\kappa _{l'}}}}} } \Big\|^2}}}=\\
 &\ \  \sum\limits_{k = 0}^{K - 1} {{{\left\| {{{\bf{V}}_k^H}{{\bf{\bar X}}}{{\bf{u}}_k}} \right\|}^2}} = \sum\limits_{k = 0}^{K - 1} {{{\left\| {{\bf{V}}_k^H{{\bf{w}}_k}} \right\|}^2}} ,
 \end{aligned}
 \end{equation}
 where ${{\bf{V}}_k} \in {{\mathbb C}^{(\sum {{{\bar r}_{l'}}}) \times {M_t} }} \triangleq [{\bf{\bar H}}_1^ \bot {e^{ - j\frac{{2\pi }}{{K}}k{\kappa _1}}}, \cdots ,$ ${\bf{\bar H}}_{L'}^ \bot {e^{ - j\frac{{2\pi }}{{K}}k{\kappa _{L'}}}}]^H$.

 As a result, problem (P1) can be equivalently written as
 \begin{equation}\label{reducedProblemVariableComb}
 \begin{aligned}
 \mathop {\max }\limits_{{\bf {\bar X}}, \left\{ {{{\bf{w}}_k}},{{{\bf{u}}_k}} \right\}_{k = 0}^{K - 1}  }& \  \frac{1}{{K}}\sum\limits_{k = 0}^{K - 1}  {{{\log }_2}\left( 1 + {{\left| {{\bf{g}}_k^H{{\bf{w}}_k}} \right|}^2} \right)}\\
 {\rm{s.t.}} &\ \ \sum\limits_{k = 0}^{K - 1} {{{\left\| {{{\bf{V}}_k^H}{{\bf{w}}_k}} \right\|}^2}}  \le KP,\\
 &\ \ {{\bf{w}}_k} = {{\bf{\bar X}}}{{\bf{u}}_k},\ \forall k \in \left[0,K-1\right].
 \end{aligned}
 \end{equation}
 It is observed from \eqref{reducedProblemVariableComb} that the frequency-domain beamforming vectors $\{ {{\bf{u}}_k}\} _{k = 0}^{K - 1}$ and the time-domain beamforming matrix ${\bf{\bar X}}$ appear in the objective function and the power constraint in the form of their product ${\bf{w}}_k = {\bf{\bar X}}{{\bf{u}}_k}$, $\forall k$. As a result, we first consider the relaxed problem of \eqref{reducedProblemVariableComb} by temporarily relaxing the last constraints, ${{\bf{w}}_k} = {\bf{\bar X}}{{\bf{u}}_k}$, $\forall k \in \left[0,K-1\right]$. The following result is useful for finding the optimal solution to the relaxed problem of \eqref{reducedProblemVariableComb}.

 \begin{theorem}\label{TheoremSubProblem1gkLieinVk}
 For each sub-carrier $k$, ${{\bf{g}}_k}$ lies in the space spanned by the columns of ${{\bf{V}}_k}$, i.e., ${{\bf{g}}_k} = {{\bf{V}}_k}{{\bf{e}}_k}$, where
 \begin{equation}
 {{\bf{e}}_k} = \frac{1}{{ \sigma }}\sum\limits_{l = 1}^L {{{\bf{h}}_l}{e^{j\frac{{2\pi }}{K}k\left( {{n_l} - {n_{\max }} + {n'_{{\rm{span}}}}} \right)}}}.
 \end{equation}
 \end{theorem}

 \begin{IEEEproof}
 Please refer to Appendix~\ref{proofOfTheoremSubProblem1gkLieinVk}.
 \end{IEEEproof}

 With Theorem \ref{TheoremSubProblem1gkLieinVk}, the relaxed problem of \eqref{reducedProblemVariableComb} can be equivalently written as
 \begin{equation}\label{reducedProblemVariableCombequivalence1}
 \begin{aligned}
 \mathop {\max }\limits_{ \left\{ {{{\bf{w}}_k}} \right\}_{k = 0}^{K - 1}  }& \  \frac{1}{{K}}\sum\limits_{k = 0}^{K - 1} {{{\log }_2}\left( {1 + {{\left| {{\bf{e}}_k^H{\bf{V}}_k^H{{\bf{w}}_k}} \right|}^2}} \right)}\\
 {\rm{s.t.}} &\ \ \sum\limits_{k = 0}^{K - 1} {{{\left\| {{{\bf{V}}_k^H}{{\bf{w}}_k}} \right\|}^2}}  \le KP.
 \end{aligned}
 \end{equation}
 Let the (reduced) singular value decomposition (SVD) of ${\bf{V}}_k^H$ be expressed as ${{\bf{V}}_k^H} = {{\bf{A}}_k}{{\bf{\Sigma }}_k}{\bf{B}}_k^H$, where ${{\bf{A}}_k} \in {\mathbb C}^{{M_t} \times {r_k}}$, ${{\bf{B}}_k} \in {\mathbb C}^{(\sum {{\bar r}_{l'}} ) \times {r_k}}$, and ${\bf \Sigma} _k \in {\mathbb C}^{{r_k} \times {r_k}}$ contains $r_k$ positive singular values of ${{\bf{V}}_k^H}$, $\forall k$. Then we have the following theorem.

 \begin{theorem}\label{generalCaseDAMOFDMOptimalSolution}
 An optimal solution to problem \eqref{reducedProblemVariableCombequivalence1} is
 \begin{equation}\label{optimalSolutionToGeneralCase}
 {{\bf{w}}_k^{\star}} = \sqrt {{\mu _k^{\star}}} {{\bf{B}}_k}{\bf{\Sigma }}_k^{ - 1}\frac{{{\bf{A}}_k^H{{\bf{e}}_k}}}{{\left\| {{\bf{A}}_k^H{{\bf{e}}_k}} \right\|}}, \ \forall k,
 \end{equation}
 where ${\mu _k^{\star}}$ denotes the optimal power allocation for sub-carrier $k$ that is given by \eqref{generalCasePowerAllocation}.
 \end{theorem}

 \begin{IEEEproof}
 Please refer to Appendix~\ref{proofOfgeneralCaseDAMOFDMOptimalSolution}.
 \end{IEEEproof}

 With the optimal solution $\{{{\bf{w}}_k^{\star}}\}_{k = 0}^{K - 1}$ to problem \eqref{reducedProblemVariableCombequivalence1} obtained in Theorem \ref{generalCaseDAMOFDMOptimalSolution}, we aim to find the solution to the original problem \eqref{reducedProblemVariableComb}. Since \eqref{reducedProblemVariableCombequivalence1} is a relaxed problem of \eqref{reducedProblemVariableComb} by discarding the constraints ${{\bf{w}}_k} = {\bf{\bar X}}{{\bf{u}}_k}$, $\forall k$, if we can find the time-domain beamforming matrix ${\bf{\bar X}}$ and the frequency-domain beamforming vectors $\{ {{\bf{u}}_k}\} _{k = 0}^{K - 1}$ such that ${\bf{\bar X}}{{\bf{u}}_k} = {\bf{w}}_k^{\star}$, $\forall k$, then the resulting ${\bf{\bar X}}$ and $\{ {{\bf{u}}_k}\} _{k = 0}^{K - 1}$ must be optimal to the original problem \eqref{reducedProblemVariableComb}, since it satisfies all constraints in \eqref{reducedProblemVariableComb} and achieves identical objective value of the relaxed problem \eqref{reducedProblemVariableCombequivalence1}. Therefore, the remaining task is to find ${\bf{\bar X }}$ and ${\bf{\bar U }}$ such that ${\bf{W}}^{\star} = {\bf{\bar X \bar U}}$, where ${{\bf{W}}^{\star}} \in {\mathbb C}^{({\sum {{\bar r}_{l'}}}) \times K } \triangleq \left[ {{{\bf{w}}_0^{\star}}, \cdots ,{{\bf{w}}_k^{\star}}, \cdots ,{{\bf{w}}_{K - 1}^{\star}}} \right]$ and ${{\bf{\bar U}}}  \in {\mathbb C}^{{M_t} \times K } \triangleq \left[ {{{\bf{u}}_0}, \cdots ,{{\bf{u}}_k}, \cdots ,{{\bf{u}}_{K - 1}}} \right]$. Depending on the relationship of $\sum {{\bar r}_{l'}}$, $K$, and $M_t$, the following three cases are respectively considered.

 \emph{Case 1}: ${M_t} \ge K$. In this case, the matrix ${\bf{W}}^{\star}$ can be expressed as
 \begin{equation}\label{matrixDecompositionCase1}
 {{\bf{W}}^{\star}} = \left[ {{{\bf{W}}^{\star}},{{\bf{0}}_{\left( {\sum {{{\bar r}_{l'}}} } \right) \times \left( {{M_t} - K} \right)}}} \right]\left[ \begin{array}{l}
 {{\bf{I}}_K}\\
 {{\bf{0}}_{\left( {{M_t} - K} \right) \times K}}
 \end{array} \right].
 \end{equation}
 Thus, one optimal solution to problem \eqref{reducedProblemVariableComb} is
 \begin{equation}\label{Case1OptimalPrecodingMatrix}
 {{\bf{\bar X}}^{\star}} = \left[ {{{\bf{W}}^{\star}},{{\bf{0}}_{\left( {\sum {{{\bar r}_{l'}}} } \right) \times \left( {{M_t} - K} \right)}}} \right],
 \end{equation}
 and
 \begin{equation}\label{Case1OptimalBeamformingVector}
 {\bf{u}}_k^{\star} = {\left[ {\bf{\bar U}}^ {\star}  \right]_{:,k + 1}} = {\left[ \begin{array}{l}
{{\bf{I}}_K}\\
{{\bf{0}}_{\left( {{M_t} - K} \right) \times K}}
\end{array} \right]_{:,k + 1}}, \ \forall k,
 \end{equation}
 where ${\left[ {{{{\bf{\bar U}}}^ {\star} }} \right]_{:,k + 1}}$ denotes the $\left(k+1\right)$th column of ${\bf{\bar U}}^ {\star}$.

 \emph{Case 2}: $\sum {{{\bar r}_{l'}}}  \le {M_t} < K$. In this case, let the SVD of ${\bf{W}}^{\star}$ be ${\bf{W}}^{\star} = {\bf{\tilde A\tilde \Sigma }}{{\bf{\tilde B}}^H}$, where ${\bf{\tilde A}} \in {{\mathbb C}^{(\sum {{{\bar r}_{l'}}} ) \times (\sum {{{\bar r}_{l'}}} )}}$, ${\bf{\tilde B}} \in {{\mathbb C}^{K \times K}}$, and
 ${\bf{\tilde \Sigma }} \in {{\mathbb C}^{(\sum {{{\bar r}_{l'}}} ) \times K}}$ contains the singular values that are arranged in descending order. Then we have
 \begin{equation}\label{matrixDecompositionCase2}
 {{\bf{W}}^{\star}} = \left[ {{\bf{\tilde A}},{{\bf{0}}_{(\sum {{{\bar r}_{l'}}} ) \times ({M_t} - (\sum {{{\bar r}_{l'}}} ))}}} \right]\left[ \begin{split}
 &\ \ \ \ {\bf{\tilde \Sigma }}{{{\bf{\tilde B}}}^H}\\
 &{{\bf{0}}_{({M_t} - (\sum {{{\bar r}_{l'}}} )) \times K}}
 \end{split} \right],
 \end{equation}
 i.e., we can also find ${\bf{\bar X}}$ and $\{ {{\bf{u}}_k}\} _{k = 0}^{K - 1}$ so that ${\bf{\bar X}}{{\bf{u}}_k} = {\bf{w}}_k^{\star}$, $\forall k$. In this case, one optimal solution to problem \eqref{reducedProblemVariableComb} is
 \begin{equation}\label{Case2OptimalPrecodingMatrix}
 {{\bf{\bar X}}^{\star}} = \left[ {{\bf{\tilde A}},{{\bf{0}}_{(\sum {{{\bar r}_{l'}}} ) \times ({M_t} - (\sum {{{\bar r}_{l'}}} ))}}} \right],
 \end{equation}
 and
 \begin{equation}\label{Case2OptimalBeamformingVector}
 {\bf{u}}_k^{\star} = {\left[ {\bf{\bar U}}^ {\star}  \right]_{:,k + 1}} = \left[ \begin{split}
 &\ \ \ \ {\bf{\tilde \Sigma }}{{{\bf{\tilde B}}}^H}\\
 &{{\bf{0}}_{({M_t} - (\sum {{{\bar r}_{l'}}} )) \times K}}
 \end{split} \right]_{:,k + 1}, \ \forall k.
 \end{equation}

 \emph{Case 3}:  ${M_t} < \min \left( {K,\sum {{{\bar r}_{l'}}} } \right)$. In this case, there is no direct way to find ${\bf{\bar X}}$ and ${\bf{\bar U}}$ such that ${\bf{W}}^{\star} = {\bf{\bar X \bar U}}$ is satisfied. Therefore, we construct an approximate solution to \eqref{reducedProblemVariableComb} as
 \begin{equation}\label{Case3OptimalPrecodingMatrix}
 {{\bf{\bar X}}^{\star}} = {\left[ {{\bf{\tilde A}}} \right]_{:,1:{M_t}}},
 \end{equation}
 where $[ {\bf{\tilde A}}]_{:,1:{M_t}}$ denotes the first $M_t$ columns of $\bf {\tilde A}$, and
 \begin{equation}\label{Case3OptimalBeamformingVector}
  {\bf{\bar U}}^{\star} = {\left[ {{\bf{\tilde \Sigma }}{{\bf{\tilde B}}^H}} \right]_{1:{M_t},:}},
 \end{equation}
 where $[ {{\bf{\tilde \Sigma }}{{\bf{\tilde B}}^H}} ]_{1:{M_t},:}$ denotes the first $M_t$ rows of ${\bf{\tilde \Sigma }}{{{\bf{\tilde B}}}^H}$.

 The proposed joint frequency- and time-domain beamforming design for DAM-OFDM is summarized in Algorithm~\ref{alg1}. Furthermore, we provide complexity analysis for the proposed joint frequency- and time-domain beamforming design. In Algorithm~\ref{alg1}, step 4 has the complexity ${\cal O}({M_t}\sum\nolimits_{l' = 1}^{L'} {{\left| {{{\cal L}}_{l'}} \right|}^2} )$ due to the computations of orthogonal complement. Step 5 has the complexity ${\cal O}(\sum\nolimits_{k = 0}^{K - 1} {(r_k^3 + {{\bar r}_{{\rm{sum}}}}r_k^2 + {M_t}{r_k})} )$, with ${{\bar r}_{{\rm{sum}}}} = \sum\nolimits_{l' = 1}^{L'} {{{\bar r}_{l'}}}$. In step 6, it is observed that only Cases 2 and 3 need to perform SVD, and thus the (worst-case) complexity is ${\cal O}(K{{\bar r}_{{\rm{sum}}}}\min (K,{{\bar r}_{{\rm{sum}}}}))$. The complexity for step 7 is ${\cal O}(M_t^2{{\bar r}_{{\rm{sum}}}})$ due to the matrix multiplication. As a result, the overall complexity of Algorithm~\ref{alg1} is ${\cal O}( {M_t}\sum\nolimits_{l' = 1}^{L'} {{{\left| {{{\rm{{\cal L}}}_{l'}}} \right|}^2}} + \sum\nolimits_{k = 0}^{K - 1} {( {r_k^3 + {{\bar r}_{\rm{sum}}}r_k^2 + {M_t}{r_k}} )}  + K{{\bar r}_{{\rm{sum}}}}\min ( {K,{{\bar r}_{{\rm{sum}}}} } ) + M_t^2{{\bar r}_{{\rm{sum}}}}  )$.

 \begin{algorithm}[t]
 \caption{Joint Frequency- and Time-Domain Beamforming Design for DAM-OFDM.}
 \label{alg1}
 \begin{algorithmic}[1]
 \STATE  \textbf{Input:} The multi-path channel $\{{{\bf{h}}_l}, {n_l}\} _{l = 1}^L$, the number of delay pre-compensations $L'$, and the desired delay spread ${n'_{\rm span}}<{n_{\rm span}}$.
 \STATE  Set the delay pre-compensations ${\kappa _{l'}} = {n_{\max }} - {n_{L - L' + l'}}$, $\forall l' \in \left[1,L'\right]$.
 \STATE For each ${l'} \in \left[1,L'\right]$, determine the multi-paths outside the desired delay window $\{ {{{\bf{\bar H}}}_{l'}}\} _{l' = 1}^{L'}$ based on \eqref{ineffectiveMultiPath}.
 \STATE Obtain the orthonormal basis ${\bf{\bar H}}_{l'}^ \bot $ for the orthogonal complement of ${{{\bf{\bar H}}}_{l'}}$.
 \STATE  Obtain the optimal solution $\{{{\bf{w}}_k^{\star}}\}_{k = 0}^{K - 1}$ to the relaxed problem \eqref{reducedProblemVariableCombequivalence1} based on Theorem \ref{generalCaseDAMOFDMOptimalSolution}.
 \STATE Obtain the solution ${{\bf{\bar X}}^{\star}}$ and $\{ {{\bf{u}}_k^{\star}}\} _{k = 0}^{K - 1}$ to problem \eqref{reducedProblemVariableComb} based on \eqref{Case1OptimalPrecodingMatrix}, \eqref{Case1OptimalBeamformingVector} and \eqref{Case2OptimalPrecodingMatrix}-\eqref{Case3OptimalBeamformingVector}, respectively.
 \STATE Obtain the optimized time-domain beamforming matrices ${\bf{F}}_{l'}^{\star} = {\bf{\bar H}}_{l'}^ \bot {\bf{\bar X}}_{l'}^{\star}$, $\forall l' \in \left[1,L'\right]$.
 \STATE \textbf{Output:} The delay pre-compensations and the
 time-domain beamforming matrices $\{ {\kappa _{l'}},{{\bf{F}}_{l'}^{\star}}\} _{l' = 1}^{L'}$, and the  frequency-domain beamforming vectors  $\{ {{\bf{u}}_k^{\star}}\} _{k = 0}^{K - 1}$.
 \end{algorithmic}
 \end{algorithm}

 Last, we study the impact of CP overhead for DAM-OFDM. Each OFDM symbol has duration $\left( {K + {N_{\rm{CP}}}} \right){T_s}$, and the number of OFDM symbols for each channel coherence time is ${n_{\rm{OFDM}}} = {n_c}/\left( {K + {N_{\rm{CP}}}} \right)$. The effective spectral efficiency by taking into account the CP overhead is then given by
 \begin{equation}\label{effectiveSEDAMBasedOFDMGeneralCase}
 \begin{aligned}
 R  &= \frac{{{n_c} - {n_{\rm{OFDM}}}{N_{\rm{CP}}}}}{{{n_c}}}\frac{1}{{K}}\sum\limits_{k = 0}^{K - 1} {{{\log }_2}\left( {1 + {\gamma}_k} \right)}\\
 &= \frac{1}{{K + {N_{{\rm{CP}}}}}}\sum\limits_{k = 0}^{K - 1} {{{\log }_2}\left( {1 + {\gamma _k}} \right)}.
 \end{aligned}
 \end{equation}
 By choosing the CP length for DAM-OFDM as $N_{\rm CP} = n'_{\rm span} < n_{\rm span}$, the CP overhead is thus ${{n_{\rm{OFDM}}}{n'_{\rm{span}}}}/{n_c} = {n'_{\rm{span}}}/({n'_{\rm{span}}}+K)$. As a comparison, for the conventional OFDM without DAM, denote by $K'$ and $N'_{\rm CP}$ the number of sub-carriers and its CP length, respectively. Similarly, by letting  $N'_{\rm CP} =n_{\rm span}$, the CP overhead for the conventional OFDM is given by $ {n_{\rm{span}}}/({n_{\rm{span}}}+K')$. It is observed that compared with conventional OFDM, thanks to the reduced channel delay spread from $n_{\rm span}$ to ${n'_{\rm{span}}}$, DAM-OFDM is able to save the CP overhead if the same number of sub-carriers $K=K'$ is used. On the other hand, DAM can also use fewer sub-carriers $K$ without increasing the CP overhead, which is desirable since it mitigates the practical issue of PAPR for OFDM. In particular, for perfect delay alignment with $n'_{\rm span} = 0$, DAM-OFDM requires no CP overhead for each OFDM symbol, i.e., CP-free DAM-OFDM transmission. In this case, only a guard interval of length $n_{\max }$ is needed  at the beginning of each coherence block, whose overhead reduces to ${n_{\max }}/{n_c}$.

\section{Simulation Results}\label{sectionNumericalResults}
 In this section, simulation results are provided to evaluate the performance of the proposed DAM technique. We consider a system with carrier frequency $f = 28$ GHz, the total bandwidth $B=128$ MHz, and the noise power spectrum density $N_0 = -174$ dBm/Hz. The transmitter is equipped with a ULA with adjacent elements separated by half-wavelength. The channel coherence time is ${T_c} = 1$ ms. Therefore, the total number of signal samples within each coherence time is ${n_c} = B{T_c }=1.28 \times {10^5}$. The number of temporal-resolvable multi-paths with non-negligible path strength is set as $L=5$, whose path delays are uniformly distributed in $\left[ {0,{\tau_{\rm{max}}}} \right]$, with $\tau_{\rm{max}} = 312.5$ ns. Therefore, the upper bound of delay spread over all channel coherence blocks is ${{\tilde n}_{{\rm{span}}}} = {{\tilde n}_{\max }} = {\tau _{\max }}B = 40$. The number of sub-paths $\mu_l$, $\forall l$, is uniformly distributed in $\left[ {1,{\mu_{\rm{max}}}} \right]$, with ${\mu_{\max}}=3$. The AoDs of all the sub-paths are randomly distributed in the interval $\left[ { - {{60}^\circ },{{60}^\circ }} \right]$. Furthermore, the complex-valued path gains ${\alpha _l}$, $\forall l$, are generated based on the model developed in \cite{akdeniz2014millimeter}. For the benchmark OFDM scheme without DAM, a CP of length ${N'_{\rm CP}}={\tilde n}_{\rm span } = 40$ is used.

 \begin{figure}[!t]
 \centering
 \centerline{\includegraphics[width=3.5in,height=2.625in]{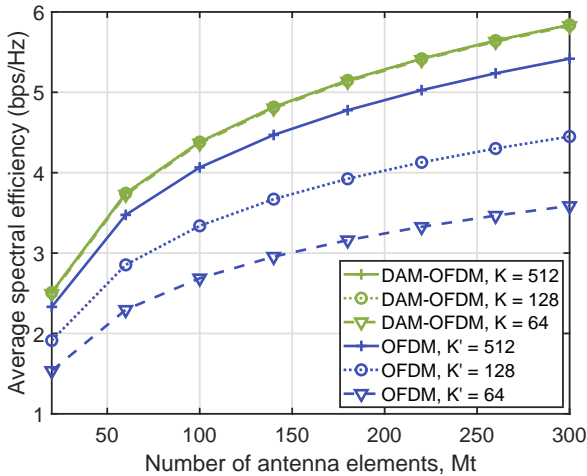}}
 \caption{Comparison of the conventional OFDM and the proposed DAM-OFDM in terms of average spectral efficiency. For DAM-OFDM, perfect delay alignment is achieved with $L'=L=5$.}
 \label{generalCaseAverageSEDifferentSubcarriers}
 \end{figure}
 Fig.~\ref{generalCaseAverageSEDifferentSubcarriers} shows the average spectral efficiency over ${10^4}$ channel realizations versus the number of antennas for the conventional OFDM and our proposed DAM-OFDM. For the latter, perfect delay alignment with ${n'_{\rm span}} = 0$ is achieved, by introducing $L'=L=5$ delay pre-compensations. The transmit power is $P=30$ dBm. It is observed from Fig.~\ref{generalCaseAverageSEDifferentSubcarriers} that the average spectral efficiency of DAM-OFDM remains almost unchanged as the number of sub-carriers $K$ varies, while the conventional OFDM incurs significant performance loss when the number of sub-carrier $K'$ decreases. This is expected due to the saving of CP overhead with our proposed DAM-OFDM. Thanks to perfect delay alignment achieved, the proposed DAM-OFDM requires no CP overhead for each OFDM symbol, but only a guard interval of length ${{\tilde n}_{\max }} = 40$ at the beginning of each coherence block. Therefore, the overhead is ${{\tilde n}_{\max }}/{n_c} = \frac{40}{{1.28 \times {{10}^5}}} = 0.031\% $, which is independent of the number of sub-carriers $K$. By contrast, for the conventional OFDM, the CP overhead critically depends on the number of sub-carriers $K'$. Specifically, for $K'=512$, the overhead is given by ${{\tilde n}_{\rm span }}/({{\tilde n}_{\rm span }} + K') = \frac{{40}}{{552}} = 7.25\% $, and it increases to $23.8\%$ and $38.5\%$ for $K'=128$ and $64$, respectively. It is also observed that the proposed DAM-OFDM achieves significantly higher spectral efficiency than the conventional OFDM, even when less sub-carriers are used. This is quite appealing since it helps mitigate the practical issue of PAPR for OFDM transmission.

 \begin{figure}[!t]
 \centering
 \centerline{\includegraphics[width=3.5in,height=2.625in]{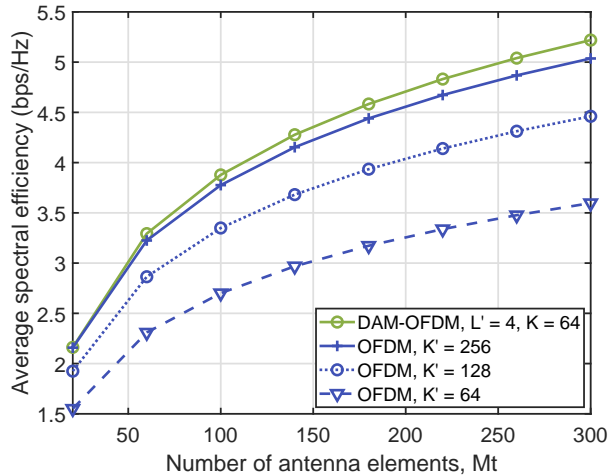}}
 \caption{Further comparison of the conventional OFDM and the proposed DAM-OFDM in terms of average spectral efficiency. For DAM-OFDM, non-perfect delay alignment with $n'_{\rm span} = 5$ and $L'= 4$ is considered.}
 \label{generalCaseAverageSEDifferentLprime}
 \end{figure}
 As a further comparison, Fig.~\ref{generalCaseAverageSEDifferentLprime} shows the average spectral efficiency for OFDM and DAM-OFDM, where DAM is applied so that the delay spread is reduced to $n'_{\rm span} = 5$, with $L' = 4$ delay pre-compensations. It is observed that even with non-perfect delay alignment, the proposed DAM-OFDM with $K= 64$ still outperforms the conventional OFDM that uses more sub-carriers.

 \begin{figure}[!t]
 \centering
 \centerline{\includegraphics[width=3.5in,height=2.625in]{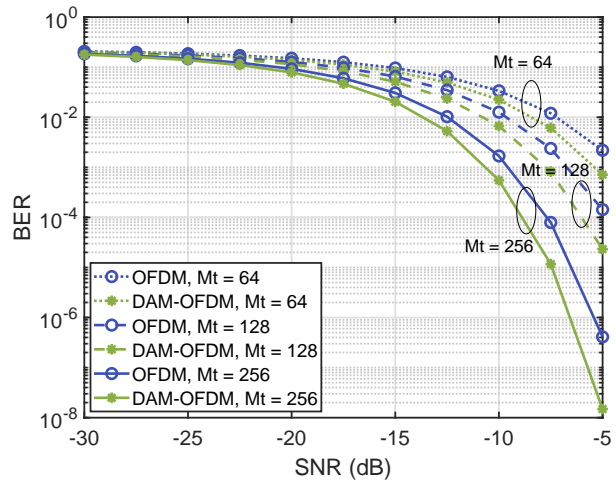}}
 \caption{BER comparison for OFDM and DAM-OFDM with different number of transmit antennas.}
 \label{BERSNRComparisionAntennaNumber}
 \end{figure}
 Fig.~\ref{BERSNRComparisionAntennaNumber} shows the BER performance versus SNR for the conventional OFDM and the proposed DAM-OFDM, by using 256 quadrature amplitude modulation (QAM) modulation. For the above frequency-selective channel setup, the SNR is defined as ${\rm{SNR}} \triangleq P{\mathbb E}\left[ \beta  \right]/{\sigma ^2}$, where $\beta$ denotes the large-scale attenuation including distance-dependent path loss and shadowing \cite{zeng2016millimeter}. For DAM-OFDM, perfect delay alignment with $n'_{\rm span} = 0$ is achieved. The number of sub-carriers is set as $K = K' = 128$. Let ${P_{e}}\left( \cdot \right)$ denote the BER expression of QAM in the AWGN channel. Then the BER of the OFDM system can be written as \cite{xia2001precoded}
 \begin{equation}\label{OFDMBER}
 {P_{e,{\rm{OFDM}}}} = \frac{1}{K'}\sum\limits_{k = 1}^{K'} {{P_e}\left( {\frac{{{K'}{p_k^{\star}}{{\left\| {{\bf{h}}\left[ k \right]} \right\|}^2}}}{{\left( {{K'} + {{\tilde n}_{{\rm{span}}}}} \right){\sigma ^2}/{K'}}}} \right)},
 \end{equation}
where $p_k^{\star}$ is the transmit power of the $k$th sub-carrier, e.g., the classic water-filling (WF) power allocation \cite{goldsmith2005wireless}. The BER performance of OFDM in Fig.~\ref{BERSNRComparisionAntennaNumber} is then obtained based on \eqref{OFDMBER}. On the other hand, thanks to perfect delay alignment, the frequency-domain effective channel for DAM-OFDM in \eqref{effectiveFrequencyDomainChannelDAMOFDM} reduces to ${{{\bf{\tilde h}}}^H}\left[ k \right] = \frac{1}{\sqrt K}\sum\nolimits_{l = 1}^L {{\bf{h}}_l^H{\bf{ H}}_l^ \bot {{{\bf{ X}}}_l}}$, $\forall k$, which implies that all the $K$ sub-carriers experience identical channel gain, i.e., ${\gamma _k} = \gamma$, $\forall k$. In this case, the BER expression of DAM-OFDM is ${P_{e,{\rm{DAM-OFDM}}}} = {{P_e}\left( {\gamma} \right)} $. It is observed from Fig.~\ref{BERSNRComparisionAntennaNumber} that as the number of transmit antennas $M_t$ increases, the BER performance of both DAM-OFDM and OFDM improves, as expected. It is also observed that for the considered setup, DAM-OFDM gives better BER performance than OFDM for the three cases of $M_t = 64,128,256$, thanks to the saving of CP overhead, as reflected by the term ${K'}/({K'} + {{\tilde n}_{\rm{span}}})$ in \eqref{OFDMBER}.

 \begin{figure}
 \centering
 \subfigure[Perfect DAM, $n'_{\rm span} = 0$]{
 \begin{minipage}[t]{0.5\textwidth}
 \centering
 \centerline{\includegraphics[width=3.5in,height=2.625in]{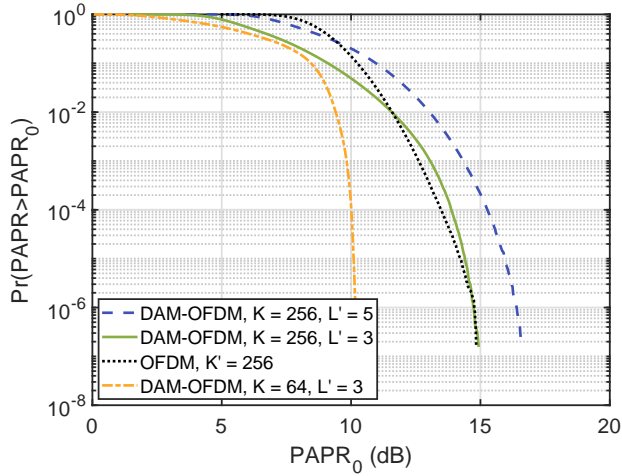}}
 \end{minipage}
 }
 \subfigure[Non-perfect DAM, $n'_{\rm span} = 5$]{
 \begin{minipage}[t]{0.5\textwidth}
 \centering
 \centerline{\includegraphics[width=3.5in,height=2.625in]{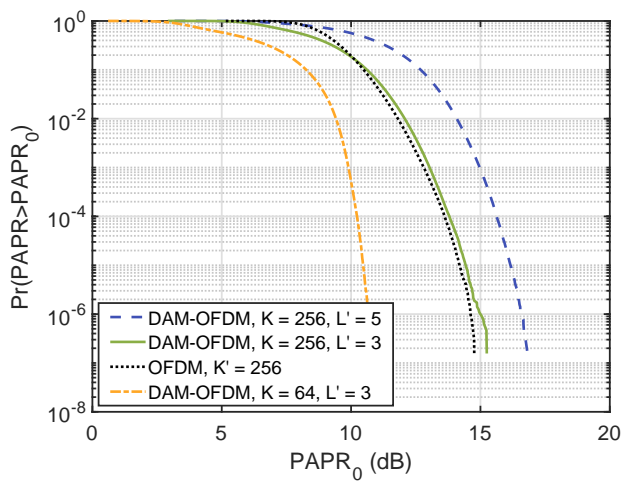}}
 \end{minipage}
 }
 \caption{PAPR comparison for the conventional OFDM and the proposed DAM-OFDM.}
 \label{PAPRComparison}
 \end{figure}

 Last, Fig.~\ref{PAPRComparison} shows the PAPR comparison for OFDM and the DAM-OFDM, with 128QAM. The PAPR performance is evaluated by the commonly used complementary cumulative distribution function (CCDF), which denotes the probability that the PAPR of a data block exceeds a given threshold \cite{han2005overview,hung2014papr}. The number of antennas is $M_t = 128$. It is observed that for the considered setup, when the same number of sub-carriers is used, DAM-OFDM has similar PAPR as OFDM for $L' = 3$ and it is slightly higher when $L' = 5$. This is expected since $L'K$ signals are superimposed at the transmitter for DAM-OFDM, as can be seen in \eqref{DAMOFDMTransmitSignalTime}, while the signals from $K'$ sub-carriers are mixed on each antenna for OFDM. Fortunately, thanks to the reduction of channel delay spread, for both perfect delay alignment with $n'_{\rm span} = 0$ and non-perfect delay alignment with $n'_{\rm span} = 5$, DAM-OFDM is able to significantly reduce the PAPR by using fewer sub-carriers, while achieving even higher spectral efficiency as can be seen from Fig.~\ref{generalCaseAverageSEDifferentLprime}. It is also observed from Fig.~\ref{PAPRComparison} that compared to the conventional OFDM, the proposed DAM-OFDM provides more flexibility for PAPR reduction, not only by using fewer sub-carriers as in conventional OFDM, but also by introducing fewer delay pre-compensations $L'$, without increasing the CP overhead. This eases the requirement of linear region of power amplifiers compared to the conventional OFDM. By combining Figs.~\ref{generalCaseAverageSEDifferentSubcarriers}-\ref{PAPRComparison}, it is concluded that the proposed DAM-OFDM outperforms the conventional OFDM in terms of spectral efficiency, BER, and PAPR, thanks to the reduction of channel delay spread achieved by DAM.

\section{Conclusion}\label{sectionConclusion}
 This paper proposed a unified DAM framework to combat time-dispersive channel for efficient single- or multi-carrier transmissions via spatial-delay processing, by exploiting the high spatial resolution brought by large antenna arrays and the multi-path sparsity of mmWave and Terahertz channels. We first showed that when the number of spatial dimensions is no smaller than that of the temporal resolvable multi-paths, perfect delay alignment can be achieved to transform the time-dispersive channel to time non-dispersive channel, without sophisticated channel equalization or multi-carrier processing. Furthermore, when perfect DAM is infeasible or undesirable, the more generic DAM technique was proposed to significantly reduce the channel delay spread. Besides, the novel DAM-OFDM technique was proposed, and the joint frequency- and temporal-domain beamforming optimization framework was introduced. Extensive simulation results were provided to demonstrate the superiority of DAM-OFDM over the conventional OFDM, in terms of spectral efficiency, BER, and PAPR. Moreover, an important future work direction is to investigate the channel estimation and performance characterization under CSI errors for DAM-OFDM.

\begin{appendices}
\section{Proof of Theorem \ref{TheoremTransmitPowerDAMOFDMSignal}}\label{proofOfTheoremTransmitPowerDAMOFDMSignal}
 With \eqref{DAMOFDMTransmitSignalTime} and \eqref{diMinustaosymbolStreamDAMOFDMSymSamNum}, the transmit power of ${\bf{\bar q}}\left[i \right]$ can be expressed as
 \begin{equation}\label{transmitPowerDAMOFDMSignalProof}
 \begin{aligned}
 &{\mathbb E}\left[ {{{\left\| {{\bf{\bar q}}\left[ i \right]} \right\|}^2}}\right] = {\mathbb E}\Big[ {{{\Big\| {\sum\limits_{l = 1}^{L'} {{{\bf{\bar H}}_{l}^ \bot {\bf{\bar X}}_{l} }{\bf{d}}\left[ {i - {\kappa _{l}}} \right]} } \Big\|^2}}} \Big]\\
 &= {\mathbb E}\Big[ {{{\Big\| {\sum\limits_{l = 1}^{L'} {{{\bf{\bar H}}_{l}^ \bot {\bf{\bar X}}_{l} }{{{\bf{\bar x}}}_{\rm{t}}}\left[ {{m_{i,l}},{n_{i,l}}} \right]} } \Big\|^2}}} \Big]\\
 &= \frac{1}{K}{\mathbb E}\Big[ {{{\Big\| {\sum\limits_{l = 1}^{L'} {\sum\limits_{k = 0}^{K - 1} {{\bf{\bar H}}_{l}^ \bot {\bf{\bar X}}_{l} } {{\bf{u}}_k}s\left[ {{m_{i,l}},k} \right]{e^{j\frac{{2\pi }}{{K}}k{n_{i,l}}}}} } \Big\|^2}}} \Big]\\
 &= \frac{1}{K}\sum\limits_{k = 0}^{K - 1}  {\Big\| {\sum\limits_{l = 1}^{L'} {{{\bf{\bar H}}_{l}^ \bot {\bf{\bar X}}_{l} }{{\bf{u}}_k}{e^{j\frac{{2\pi }}{{K}}k{n_{i,l}}}}} } \Big\|^2},
 \end{aligned}
 \vspace{-0.1cm}
 \end{equation}
 where ${m_{i,l}} = \left\lfloor {\frac{{i - {\kappa _{l}} + {N_{\rm{CP}}}}}{{K + {N_{\rm{CP}}}}}} \right\rfloor$ and ${n_{i,l}} = {\rm{mod}}\left( {i - {\kappa _l} + {N_{\rm{CP}}},} \right.$ $\left. {K + {N_{\rm{CP}}}} \right) - {N_{\rm{CP}}}$ denote the OFDM symbol and sample number for any given index $i$ and ${\kappa}_l$, respectively, and the last equality holds since $s\left[ {{m_{i,l}},k} \right]$ is independent across different $k$. Since the index $i = m\left( {K + {N_{\rm{CP}}}} \right) + n$, as defined below \eqref{newSymbolStreamDAMOFDM}, we have ${n_{i,l}} = {\rm{mod}}\left( {m\left( {K + {N_{{\rm{CP}}}}} \right) + n - {\kappa _l} + {N_{{\rm{CP}}}},} \right.$ $\left. {K + {N_{{\rm{CP}}}}} \right) - {N_{{\rm{CP}}}}$. Depending on whether $ n - {\kappa _{l}} + {N_{\rm{CP}}} \ge 0$ or not, we have the following two cases:.

 \emph{Case 1}: $ n - {\kappa _{l}} + {N_{\rm{CP}}} \ge 0$. In this case, ${n_{i,l}} =  n - {\kappa _{l}}$. Thus, we have
 \begin{equation}
 \begin{aligned}
 {\mathbb E}\left[ {{{\left\| {{\bf{\bar q}}\left[ i \right]} \right\|}^2}} \right] &= \frac{1}{K}\sum\limits_{k = 0}^{K - 1} {\Big\| {\sum\limits_{l = 1}^{L'} {{{\bf{\bar H}}_{l}^ \bot {\bf{\bar X}}_{l} }{{\bf{u}}_k}{e^{j\frac{{2\pi }}{K}k\left( {n - {\kappa _{l}}} \right)}}} } \Big\|^2}\\
 &= \frac{1}{K}\sum\limits_{k = 0}^{K - 1}  {\Big\| {{e^{j\frac{{2\pi }}{{K}}kn}}\sum\limits_{l = 1}^{L'} {{{\bf{\bar H}}_{l}^ \bot {\bf{\bar X}}_{l} }{{\bf{u}}_k}{e^{ - j\frac{{2\pi }}{{K}}k{\kappa _{l}}}}} } \Big\|^2}\\
 &= \frac{1}{K}\sum\limits_{k = 0}^{K - 1} {\Big\| {\sum\limits_{l = 1}^{L'} {{{\bf{\bar H}}_{l}^ \bot {\bf{\bar X}}_{l} }{{\bf{u}}_k}{e^{ - j\frac{{2\pi }}{K}k{\kappa _{l}}}}} } \Big\|^2}.
 \end{aligned}
 \vspace{-0.1cm}
 \end{equation}

 \emph{Case 2}: $ n - {\kappa _{l}} + {N_{\rm{CP}}} < 0$. In this case, ${n_{i,l}} = n - {\kappa _{l}} + K + {N_{\rm{CP}}}$. Thus, the transmit power is
 \begin{equation}
 \begin{aligned}
 {\mathbb E}\left[ {{{\left\| {{\bf{\bar q}}\left[ i \right]} \right\|}^2}}\right]  &= \frac{1}{K} \sum\limits_{k = 0}^{K - 1} {\Big\| {\sum\limits_{l = 1}^{L'} {{{\bf{\bar H}}_{l}^ \bot {\bf{\bar X}}_{l} }{{\bf{u}}_k}{e^{j\frac{{2\pi }}{{K}}k\left( {n - {\kappa _{l}} + K + {N_{\rm{CP}}}} \right)}}} } \Big\|^2}\\
 &= \frac{1}{K} \sum\limits_{k = 0}^{K - 1} {\Big\| {\sum\limits_{l = 1}^{L'} {{{\bf{\bar H}}_{l}^ \bot {\bf{\bar X}}_{l} }{{\bf{u}}_k}{e^{ - j\frac{{2\pi }}{{K}}k{\kappa _{l}}}}} } \Big\|^2}.
 \end{aligned}
 \end{equation}
 By combining the above two cases, Theorem \ref{TheoremTransmitPowerDAMOFDMSignal} is proved.
 \vspace{-0.1cm}
\section{Proof of Theorem \ref{TheoremSubProblem1gkLieinVk}}\label{proofOfTheoremSubProblem1gkLieinVk}
 By multiplying ${\bf{V}}_k$ with ${\bf{e}}_k$, we have
 \begin{equation}
 {\small
 \begin{aligned}
 {{\bf{V}}_k}{{\bf{e}}_k} = \left[ \begin{aligned}
 &{\left( {{\bf{\bar H}}_1^ \bot } \right)^H}{{\bf{e}}_k}{e^{j\frac{{2\pi }}{{K}}k{\kappa _1}}}\\
 &\ \ \ \ \ \ \ \ \ \ \ \ \ \ \vdots \\
 &{\left( {{\bf{\bar H}}_{l'}^ \bot } \right)^H}{{\bf{e}}_k}{e^{j\frac{{2\pi }}{{K}}k{\kappa _{l'}}}}\\
 &\ \ \ \ \ \ \ \ \ \ \ \ \ \ \vdots \\
 &{\left( {{\bf{\bar H}}_{L'}^ \bot } \right)^H}{{\bf{e}}_k}{e^{j\frac{{2\pi }}{{K}}k{\kappa _{L'}}}}
 \end{aligned} \right],
 \end{aligned}}
 \end{equation}
 where the $l'$th block of size ${{\bar r}_{l'}} \times 1$ can be given by
 \begin{equation}\label{thelprimeOfgk}
 \begin{aligned}
 &{\left( {{\bf{\bar H}}_{l'}^ \bot } \right)^H}{{\bf{e}}_k}{e^{j\frac{{2\pi }}{{K}}k{\kappa _{l'}}}}\\
 &= {\left( {{\bf{\bar H}}_{l'}^ \bot } \right)^H}{\frac{1}{{ \sigma }}}\sum\limits_{l = 1}^L {{{{\bf{ h}}}_l}{e^{j\frac{{2\pi }}{{K}}k\left( {{n_l} - {n_{\max }} + {n'_{\rm{span}}} + {\kappa _{l'}}} \right)}}} \\
 &\mathop  = \limits^{\left( a \right)} {\left( {{\bf{\bar H}}_{l'}^ \bot } \right)^H}{\frac{1}{{ \sigma }}}\sum\limits_{l \in {{\bar {\cal L} }_{l'}}} {{{{\bf{ h}}}_l}{e^{j\frac{{2\pi }}{{K}}k\left( {{n_l} - {n_{\max }} + {n'_{\rm{span}}} + {\kappa _{l'}}} \right)}}}\\
 &\mathop  = \limits^{\left( b \right)} {\left( {{\bf{\bar H}}_{l'}^ \bot } \right)^H}{\frac{1}{{ \sigma }}}\Big( {\sum\limits_{t = 0}^{{n'_{\rm{span}}}} {{{\bf{g}}_{l'}}\left[ t \right]{e^{j\frac{{2\pi }}{{K}}kt}}} } \Big) = {\left( {{\bf{\bar H}}_{l'}^ \bot } \right)^H}{{\bf{g}}_{kl'}},
 \end{aligned}
 \end{equation}
 where ${\left( a \right)}$ holds since ${{\bf{\bar H}}_{l'}^ \bot }$ is the orthogonal complement of ${{{\bf{\bar H}}}_{l'}}$, i.e., the multiplication of $({{\bf{\bar H}}_{l'}^ \bot })^H$ with the ${{{{\bf{ h}}}_l}}$ outside the desired delay window (i.e., $l \in {{\cal L}_{l'}}$) are zero vector for any given $l'$, and ${\left( b \right)}$ holds since the ${{{{\bf{\hat h}}}_l}}$ inside the window can be expressed as $\sum\nolimits_{t = 0}^{{n'_{\rm{span}}}} {{{\bf{g}}_{l'}}\left[ t \right]}$, together with the relationship ${n_l} + {\kappa _{l'}} = t + {n_{\max }} - {n'_{{\rm{span}}}}$, as defined in \eqref{newDefinitionEffectiveChannelVector}. It is observed that \eqref{thelprimeOfgk} is exactly the $l'$th block of ${{\bf{g}}_k}$ defined below \eqref{frequencyDomainSNRDAMOFDM}. By considering all the $L'$ blocks based on \eqref{thelprimeOfgk}, we have ${{\bf{g}}_k} = {{\bf{V}}_k}{{\bf{e}}_k}$. The proof of Theorem \ref{TheoremSubProblem1gkLieinVk} is thus completed.
 \section{Proof of Theorem  \ref{generalCaseDAMOFDMOptimalSolution}}\label{proofOfgeneralCaseDAMOFDMOptimalSolution}
 With the reduced SVD of ${{\bf{V}}_k^H} = {{\bf{A}}_k}{{\bf{\Sigma }}_k}{\bf{B}}_k^H$, let ${\bf{B}}_k^ \bot  \in {{\mathbb C}^{(\sum {{{\bar r}_{l'}}} ) \times ((\sum {{{\bar r}_{l'}}})  - {r_k})}}$ denote the orthogonal complement of ${\bf {B}}_k$, i.e., $[{{\bf{B}}_k},{\bf{B}}_k^ \bot ]$ forms an orthonormal basis for the $\left(\sum {{{\bar r}_{l'}}}\right)$-dimensional space. Then the optimization variable ${{\bf{w}}_k}$
 can be expressed as ${{\bf{w}}_k} = {{\bf{B}}_k}{{\bf{c}}_k} + {\bf{B}}_k^ \bot {\bf{c}}_k^ \bot $, where ${{\bf{c}}_k} \in {{\mathbb C}^{{r_k} \times 1}}$, and ${\bf{c}}_k^ \bot  \in {{\mathbb C}^{((\sum {{{\bar r}_{l'}}})  - {r_k}) \times 1}}$. It then follows that
 \begin{equation}
 \begin{aligned}
 {\bf{V}}_k^H{{\bf{w}}_k} &= {{\bf{A}}_k}{{\bf{\Sigma }}_k}{\bf{B}}_k^H\left[ {{{\bf{B}}_k},{\bf{B}}_k^ \bot } \right]\left[ \begin{array}{l}
 {{\bf{c}}_k}\\
 {\bf{c}}_k^ \bot
 \end{array} \right]\\
 &= {{\bf{A}}_k}{{\bf{\Sigma }}_k}\left[ {{{\bf{I}}_{{r_k}}},{\bf{0}}} \right]\left[ \begin{array}{l}
 {{\bf{c}}_k}\\
 {\bf{c}}_k^ \bot
 \end{array} \right] = {{\bf{A}}_k}{{\bf{\Sigma }}_k}{{\bf{c}}_k},
 \end{aligned}
 \end{equation}
 and ${\left\| {{\bf{V}}_k^H{{\bf{w}}_k}} \right\|^2} = {\bf{c}}_k^H{\bf{\Sigma }}_k^H{\bf{A}}_k^H{{\bf{A}}_k}{{\bf{\Sigma }}_k}{{\bf{c}}_k} = {\left\| {{{\bf{\Sigma }}_k}{{\bf{c}}_k}} \right\|^2}$. Let ${{{\bf{\bar e}}}_k} \in {\mathbb C}^{{r_k} \times 1} \triangleq {\bf{A}}_k^H{{\bf{e}}_k}$, and ${{\bf{\bar c}}_k} \in {\mathbb C}^{{r_k} \times 1} \triangleq {{\bf{\Sigma }}_k}{{\bf{c}}_k}$, problem \eqref{reducedProblemVariableCombequivalence1} reduces to
 \begin{equation}\label{reducedProblemVariableCombequivalence2}
 \begin{aligned}
 \mathop {\max }\limits_{\{ {\bf{\bar c}}_k\} _{k = 0}^{K - 1} } & \ \frac{1}{K}\sum\limits_{k = 0}^{K - 1} {\log _2}\left( {1 + {{\left|  {\bf{\bar e}}_k^H{{\bf{\bar c}}_k}\right|}^2}} \right)   \\
 {\rm{s.t.}} &\ \ \sum\limits_{k = 0}^{K - 1} {{{\left\| {{\bf{\bar c}}_k} \right\|}^2}}  \le KP.
 \end{aligned}
 \end{equation}
 It immediately follows that the MRT beamforming is optimal for \eqref{reducedProblemVariableCombequivalence2}, i.e., ${{{\bf{\bar c}}}_k} = \sqrt {{\mu _k}} {{{\bf{\bar e}}}_k}/\left\| {{{{\bf{\bar e}}}_k}} \right\|$, where ${\mu}_k$ denotes the power of the $k$th sub-carrier. As such, problem \eqref{reducedProblemVariableCombequivalence2} reduces to finding the optimal power allocation via solving
 \begin{equation}\label{reducedProblemVariableCombequivalence3}
 \begin{aligned}
 \mathop {\max }\limits_{\{ {\mu}_k\} _{k = 0}^{K - 1} } & \ \frac{1}{{K}}\sum\limits_{k = 0}^{K - 1} {{{\log }_2}} \left( {1 + {\mu _k}{{\left\| {{{{\bf{\bar e}}}_k}} \right\|}^2}} \right)\\
 {\rm{s.t.}} &\ \ \sum\limits_{k = 0}^{K - 1} {{\mu _k}}  \le KP.
 \end{aligned}
 \end{equation}
 The optimal solution to the above problem can be obtained by the classic WF power allocation \cite{goldsmith2005wireless}, which is given by
 \begin{equation}\label{generalCasePowerAllocation}
 \mu _k^{\star} = {\left[ {\frac{1}{\nu } - \frac{1}{{{\left\| {{{\bf{\bar e}}_k}} \right\|}^2}}} \right]^ + }, \ \forall k,
 \end{equation}
 where ${\left[ x \right]^ + } \triangleq \max \left( {x,0} \right)$, and $1/{\nu}$ is the water-level so that the power constraint is satisfied with equality. Then the optimal beamforming vectors is given by ${{\bf{w}}_k^{\star}} = \sqrt {{\mu _k^{\star}}} {{\bf{B}}_k}{\bf{\Sigma }}_k^{ - 1}{\bf{A}}_k^H{{\bf{e}}_k}/\left\| {{\bf{A}}_k^H{{\bf{e}}_k}} \right\| + {\bf{B}}_k^ \bot {\bf{c}}_k^ \bot$, where ${\bf{c}}_k^ \bot$ is an arbitrary vector. Without loss of optimality, we set ${\bf{c}}_k^ \bot = \bf{0}$, and the optimal beamforming vectors reduce to \eqref{optimalSolutionToGeneralCase}. This thus completes the proof of Theorem \ref{generalCaseDAMOFDMOptimalSolution}.
\end{appendices}


\bibliographystyle{IEEEtran}
\bibliography{refDAM}
\end{document}